\documentclass[twocolumn]{aastex631}
\usepackage{CJK}

\newcommand{\sn}{SN\,2020jgb}

\newcommand{\tfl}{$t_\mathrm{fl}$}
\newcommand{\Mch}{$M_\mathrm{Ch}$}
\newcommand{\kms}{$\mathrm{km}\,\mathrm{s}^{-1}$}
\newcommand{\Ni}{$^{56}\mathrm{Ni}$}
\newcommand{\Msun}{\mathrm{M_\odot}}

\shorttitle{\sn}
\shortauthors{Liu et al.}
\graphicspath{{./}{figures/}}

\begin{document}
\begin{CJK*}{UTF8}{gbsn}

\title{\sn: A Peculiar Type Ia Supernova Triggered by a Helium-Shell Detonation in a Star-Forming Galaxy}

\author[0000-0002-7866-4531]{Chang~Liu (刘畅)}
\affil{Department of Physics and Astronomy, Northwestern University, 2145 Sheridan Rd, Evanston, IL 60208, USA}
\affil{Center for Interdisciplinary Exploration and Research in Astrophysics (CIERA), Northwestern University, 1800 Sherman Ave, Evanston, IL 60201, USA}

\author[0000-0001-9515-478X]{Adam~A.~Miller}
\affil{Department of Physics and Astronomy, Northwestern University, 2145 Sheridan Rd, Evanston, IL 60208, USA}
\affil{Center for Interdisciplinary Exploration and Research in Astrophysics (CIERA), Northwestern University, 1800 Sherman Ave, Evanston, IL 60201, USA}

\author[0000-0002-1633-6495]{Abigail~Polin}
\affil{The Observatories of the Carnegie Institution for Science, 813 Santa Barbara Street, Pasadena, CA 91101, USA}
\affil{TAPIR, Walter Burke Institute for Theoretical Physics, 350-17, Caltech, Pasadena, CA 91125, USA}

\author[0000-0002-2028-9329]{Anya~E.~Nugent}
\affil{Department of Physics and Astronomy, Northwestern University, 2145 Sheridan Rd, Evanston, IL 60208, USA}
\affil{Center for Interdisciplinary Exploration and Research in Astrophysics (CIERA), Northwestern University, 1800 Sherman Ave, Evanston, IL 60201, USA}

\author[0000-0002-8989-0542]{Kishalay~De}
\altaffiliation{NASA Einstein Fellow}
\affil{MIT-Kavli Institute for Astrophysics and Space Research, 77 Massachusetts Ave., Cambridge, MA 02139, USA}

\author[0000-0002-3389-0586]{Peter~E.~Nugent}
\affil{Department of Astronomy, University of California, Berkeley, CA 94720-3411, USA}
\affil{Lawrence Berkeley National Laboratory, 1 Cyclotron Road, Berkeley, CA, 94720, USA}

\author[0000-0001-6797-1889]{Steve~Schulze}
\affil{The Oskar Klein Centre, Department of Physics, Stockholm University, Albanova University Center, SE 106 91 Stockholm, Sweden}

\author[0000-0002-3653-5598]{Avishay~Gal-Yam}
\affil{Department of particle physics and astrophysics, Weizmann Institute of Science, 76100 Rehovot, Israel}

\author[0000-0002-4223-103X]{Christoffer~Fremling}
\affiliation{Caltech Optical Observatories, California Institute of Technology, Pasadena, CA 91125, USA}
\affiliation{Division of Physics, Mathematics, and Astronomy, California Institute of Technology, Pasadena, CA 91125, USA}

\author[0000-0003-3768-7515]{Shreya~Anand}
\affil{Cahill Center for Astrophysics, California Institute of Technology, Pasadena CA 91125, USA}

\author[0000-0002-8977-1498]{Igor~Andreoni}
\altaffiliation{Neil Gehrels Fellow}
\affil{Joint Space-Science Institute, University of Maryland, College Park, MD 20742, USA.}
\affil{Department of Astronomy, University of Maryland, College Park, MD 20742, USA.}
\affil{Astrophysics Science Division, NASA Goddard Space Flight Center, Mail Code 661, Greenbelt, MD 20771, USA}

\author[0000-0003-0526-2248]{Peter~Blanchard}
\affil{Center for Interdisciplinary Exploration and Research in Astrophysics (CIERA), Northwestern University, 1800 Sherman Ave, Evanston, IL 60201, USA}

\author[0000-0001-5955-2502]{Thomas~G.~Brink}
\affil{Department of Astronomy, University of California, Berkeley, CA 94720-3411, USA}

\author[0000-0002-2376-6979]{Suhail~Dhawan}
\affil{Institute of Astronomy and Kavli Institute for Cosmology, University of Cambridge, Madingley Road, Cambridge CB3 0HA, UK}

\author[0000-0003-3460-0103]{Alexei~V.~Filippenko}
\affil{Department of Astronomy, University of California, Berkeley, CA 94720-3411, USA}

\author[0000-0002-9770-3508]{Kate~Maguire}
\affil{School of Physics, Trinity College Dublin, The University of Dublin, Dublin 2, Ireland}

\author[0000-0001-8948-3456]{Tassilo~Schweyer}
\affil{The Oskar Klein Centre, Department of Astronomy, Stockholm University, AlbaNova, SE-106 91 Stockholm, Sweden}

\author[0000-0001-8023-4912]{Huei~Sears}
\affil{Department of Physics and Astronomy, Northwestern University, 2145 Sheridan Rd, Evanston, IL 60208, USA}
\affil{Center for Interdisciplinary Exploration and Research in Astrophysics (CIERA), Northwestern University, 1800 Sherman Ave, Evanston, IL 60201, USA}

\author[0000-0003-4531-1745]{Yashvi~Sharma}
\affil{Division of Physics, Mathematics, and Astronomy, California Institute of Technology, Pasadena, CA 91125, USA}

\author[0000-0002-3168-0139]{Matthew~J.~Graham}
\affiliation{Division of Physics, Mathematics, and Astronomy, California Institute of Technology, Pasadena, CA 91125, USA}

\author[0000-0001-5668-3507]{Steven~L.~Groom}
\affiliation{IPAC, California Institute of Technology, 1200 E. California Blvd, Pasadena, CA 91125, USA}

\author{David~Hale}
\affiliation{Caltech Optical Observatories, California Institute of Technology, Pasadena, CA 91125, USA}

\author[0000-0002-5619-4938]{Mansi~M.~Kasliwal}
\affil{Division of Physics, Mathematics, and Astronomy, California Institute of Technology, Pasadena, CA 91125, USA}

\author[0000-0002-8532-9395]{Frank~J.~Masci}
\affiliation{IPAC, California Institute of Technology, 1200 E. California Blvd, Pasadena, CA 91125, USA}

\author[0000-0003-1227-3738]{Josiah~Purdum}
\affiliation{Caltech Optical Observatories, California Institute of Technology, Pasadena, CA 91125, USA}

\author[0000-0001-8861-3052]{Benjamin~Racine}
\affiliation{Aix Marseille Univ, CNRS/IN2P3, CPPM, Marseille, France}

\author[0000-0003-1546-6615]{Jesper~Sollerman}
\affiliation{The Oskar Klein Centre, Department of Astronomy, Stockholm University, AlbaNova, SE-106 91 Stockholm, Sweden}

\author[0000-0001-5390-8563]{Shrinivas~R.~Kulkarni}
\affiliation{Division of Physics, Mathematics, and Astronomy, California Institute of Technology, Pasadena, CA 91125, USA}

\begin{abstract} 
The detonation of a thin ($\lesssim$0.03\,$\mathrm{M_\odot}$) helium shell (He-shell) atop a $\sim$$1\,\mathrm{M_\odot}$ white dwarf (WD) is a promising mechanism to explain normal Type Ia supernovae (SNe\,Ia), while thicker He-shells and less massive WDs may explain some recently observed peculiar SNe\,Ia. We present observations of \sn, a peculiar SN\,Ia discovered by the Zwicky Transient Facility (ZTF). Near maximum brightness, \sn\ is slightly subluminous (ZTF $g$-band absolute magnitude $-18.7\,\mathrm{mag} \lesssim M_g \lesssim -18.2\,\mathrm{mag}$ depending on the amount of host-galaxy extinction) and shows an unusually red color ($0.2\,\mathrm{mag}\lesssim g_\mathrm{ZTF}-r_\mathrm{ZTF}\lesssim0.4\,\mathrm{mag}$) due to strong line-blanketing blueward of $\sim$5000\,\AA. These properties resemble those of SN\,2018byg, a peculiar SN\,Ia consistent with an He-shell double detonation (DDet) SN. Using detailed radiative transfer models, we show that the optical spectroscopic and photometric evolution of \sn\ is broadly consistent with a $\sim$0.95--1.00\,$\Msun$ (C/O core + He-shell) progenitor ignited by a {$\gtrsim$0.1\,$\Msun$ He-shell. However, one-dimensional radiative transfer models without non-local-thermodynamic-equilibrium treatment cannot accurately characterize the line-blanketing features, making the actual shell mass uncertain.} We detect a prominent absorption feature at $\sim$1\,\micron\ in the near-infrared (NIR) spectrum of \sn, which {might} originate from unburnt helium in the outermost ejecta. While the sample size is limited, {we find similar 1\,\micron\ features in all the peculiar} He-shell DDet candidates with NIR spectra obtained to date. \sn\ is also the first {peculiar} He-shell DDet SN discovered in a star-forming dwarf galaxy, indisputably showing that He-shell DDet SNe occur in both star-forming and passive galaxies, consistent with the normal SN\,Ia population.
\end{abstract}


\section{Introduction} \label{sec:intro}
It has been clear for decades that Type Ia supernovae (SNe\,Ia) are caused by the thermonuclear explosions of carbon--oxygen (C/O) white dwarfs (WDs) in binary systems \citep[see][for a review]{Maoz_2014}. Nevertheless, the nature of the binary companion, as well as how it ignites the WD, remains highly uncertain. 

The helium-shell (He-shell) double detonation (DDet) scenario is one of the most promising channels to produce SNe\,Ia. In this scenario, the WD accretes from a companion to develop a helium-rich shell, which, after becoming sufficiently massive, could detonate. Such a detonation sends a shock wave into the C/O core to trigger a runaway thermonuclear explosion that inevitably disrupts and destroys the entire WD \citep{Nomoto_1982a, Nomoto_1982b, Woosley_1986, Livne_1990, Woosley_1994, Livne_1995}. This DDet mechanism can produce explosions of WDs below the Chandrasekhar-mass (\Mch).

There are several observational benchmarks for He-shell DDet SNe. Shortly after the ignition of the He-shell, the decay of radioactive material in the helium ashes may power a detectable flash \citep{Woosley_1994,Fink_DD_2010,Kromer_DD_2010}. The Fe-group elements in the ashes will blanket blue photons with wavelengths $\lesssim$5000\,\AA\ \citep{Kromer_DD_2010}, the duration of which depends on the mass of the He-shell. For shells that are sufficiently thick, \citet{Boyle2017_Helium} suggest that the unburnt helium could provide an observational signal in near-infrared (NIR) spectra, and for those with a low progenitor mass ($\lesssim$1.0\,$\Msun$), \citet{polin_nebular_2021} predict significant [\ion{Ca}{2}] emission in the nebular phase of the SNe.

The He-shell DDet scenario could naturally account for the observational diversity in the SN\,Ia population. Using different sets of He-shell mass and C/O core mass, one can reproduce a variety of observables in ``normal'' SNe\,Ia with typical luminosities and spectral features near maximum brightness \citep[e.g.,][]{Townsley_2019,Magee_2021,Shen_2D_2021}, or peculiar subluminous ones \citep[e.g.,][]{polin_observational_2019}. 

For the He-shell DDet SNe that show ``normal'' characteristics near peak brightness, the mass of the C/O core should be $\gtrsim$1\,$\mathrm{M_\odot}$, and the mass of the He-shell is expected to be low \citep[$\lesssim$0.03\,$\mathrm{M_\odot}$;][]{polin_observational_2019,Magee_2021,Shen_2D_2021}. Recently, it was reported that SN\,2018aoz \citep{Ni_2022}, an SN\,Ia showing a rapid redward color evolution within $\sim$12\,hr after first light, could be explained by a sub-\Mch\ DDet model (a 1.05\,$\mathrm{M_\odot}$ C/O core and a 0.01\,$\mathrm{M_\odot}$ He-shell). After this red excess, the photometric evolution is consistent with that of normal SNe\,Ia, when the ashes of the He-shell become optically thin. However, some of its properties at maximum light and in the nebular phase are not consistent with an He-shell DDet scenario \citep{Ni_2022b}, making its nature debatable. 

To date, only a small fraction of SNe\,Ia have been discovered sufficiently early for possible detection of {a flux excess} \citep[e.g.,][]{Deckers_2022}, {which has been identified in a handful of SNe\,Ia such as SN\,2012cg \citep{Marion_12cg_2016}, iPTF14atg \citep{Cao_2015}, SN\,2016jhr \citep{jiang_16jhr_2017}, SN\,2017cbv \citep{Hosseinzadeh_17cbv_2017}, SN\,2018oh \citep{Dimitriadis_18oh_2019}, SN\,2019yvq \citep{Miller_2020}, SN\,2020hvf \citep{Jiang_20hvf_2021}, and SN\,2021aefx \citep{Ashall_21aefx_2022,Hosseinzadeh_21aefx_2022}.} While there could be a large underlying population of normal SNe\,Ia triggered by He-shell DDet, currently it is hard to verify this scenario. 

In contrast, if the He-shell is sufficiently massive, such that the ashes of the shell remain optically thick over a much more extended time, the SN could appear unusually red even near maximum light. Such peculiar SNe\,Ia could be normal in brightness: SN\,2016jhr is the only reported event that shows a normal peak luminosity ($M_B\approx-18.8$\,mag), but it exhibits an early red flash and maintains a red $g-r$ color throughout its evolution \citep{jiang_16jhr_2017}. Its photometric evolution as well as its spectrum around maximum could be explained by a near-\Mch\ DDet model. 
WD explosions with a total progenitor mass $<$1\,$\mathrm{M_\odot}$ are expected to be subluminous. SN\,2018byg \citep{de_18byg_2019} is a prototype of this subclass. {Other candidates include OGLE-2013-SN-079 \citep{Inserra_OGLE13_079_2015}, SN\,2016dsg \citep{Dong_16dsg_2022}, SN\,2016hnk (\citealp{jacobson-galan_16hnk_2020}; see, \citealp{galbany_16hnk_2019}), and SN\,2019ofm \citep{de_Ca_rich_2020}}. These events are faint, red, and show strong line-blanketing in spectra at maximum light. A tentative detection of unburnt helium in SN\,2016dsg was also reported by \citet{Dong_16dsg_2022}. We refer to these events as {peculiar} He-shell DDet SNe.\footnote{{In the literature they are also referred to as thick He-shell DDet SNe, which describes the physics leading to their peculiar evolution. This definition is imprecise, however, because the threshold for an He-shell to be ``thick'' depends on core mass; low-mass WDs ($\lesssim$0.8\,$\Msun$) with low-mass He-shells ($\lesssim$0.03\,$\Msun$) can still produce red, subluminous events \citep[e.g.,][]{Shen_2D_2021}.}}
The small sample size to date suggests this SN\,Ia subclass might be intrinsically rare.

In this paper, we present observations of another peculiar He-shell DDet event, \sn. This peculiar SN\,Ia highly resembles SN\,2018byg in photometric and spectroscopic properties, and exhibits a remarkable feature in the NIR spectrum that could be attributed to unburnt helium. In Section~\ref{sec:obs}, we report the observations of \sn, which are analyzed in Section~\ref{sec:analysis}, where we show its similarities with other He-shell DDet SNe and discuss the tentative \ion{He}{1} absorption features. We use a grid of He-shell DDet models to fit the data of \sn, and present the results in Section~\ref{sec:model}. Then we expand our discussion to other He-shell DDet SNe, discussing the possibly ubiquitous absorption features in their NIR spectra near 1\,\micron\ (Section~\ref{sec:1um}) and their diversity in host environments (Section~\ref{sec:host}). We draw our conclusions in Section~\ref{sec:conclusion}.

Along with this paper, we have released the data utilized in this study and the software used for data analysis and visualization. They are available online at \url{https://github.com/slowdivePTG/SN2020jgb}.

\section{Observations} \label{sec:obs}
\subsection{Discovery}

\sn\ was first discovered by the Zwicky Transient Facility \citep[ZTF;][]{Bellm_ZTF_2019a,Graham_ZTF_2019, Dekany_ZTF_2020} on 2020 May 03.463 (UT dates are used throughout this paper; MJD 58972.463) with the 48-inch Samuel Oschin Telescope (P48) at Palomar Observatory. The automated ZTF discovery pipeline \citep{Masci_ZTF_2019} detected \sn\ using the image-differencing technique of \citet{Zackay_imagesub_2016}. The candidate passed internal thresholds \citep[e.g.,][]{Duev_ZTFML_2019, Mahabal_ZTFML_2019}, leading to the production and dissemination of a real-time alert \citep{Patterson_ZTFalert_2019} and the internal designation ZTF20aayhacx. It was detected with $g_\mathrm{ZTF} = 19.86 \pm 0.15\,$mag at $\alpha_\mathrm{J2000}=17^\mathrm{h}53^\mathrm{m}12.^\mathrm{s}651$, $\delta_\mathrm{J2000}=-00^\circ51'21\farcs{81}$ and announced to the public by \citet{Fremling_report_2020}. The host galaxy, PSO J175312.663+005122.078, is a dwarf galaxy, to which \sn\ has a projected offset of only $0\farcs3$. The last nondetection limits the brightness to $r_\mathrm{ZTF} > 20.7$\,mag on 2020 April 27.477 (MJD 58966.477; 5.99\,days before the first detection). This transient was classified as an SN\,Ia by \citet{TNS_2020}. We confirm this classification via \texttt{SuperNova IDentification} \citep[\texttt{SNID};][]{Blondin_SNID_2007}, which shows \sn\ is most consistent with SNe\,Ia. Templates of other hydrogen-poor SNe, including Type Ib and Type Ic SNe, do not match the spectral sequence of \sn.

\subsection{Host-galaxy Observations}
On 2022 March 31, two years after the transient faded, we took a spectrum of its host galaxy using the DEep Imaging Multi-Object Spectrograph \citep[DEIMOS;][]{DEIMOS_2003} on the Keck-II 10\,m telescope, with a total integration time of 3200\,s. It was reduced with the \texttt{PypeIt} Python package \citep{pypeit:joss_pub}. The host exhibits strong, narrow emission lines including H$\alpha$, H$\beta$, [\ion{N}{2}] $\lambda\lambda$6548, 6583, [\ion{O}{3}] $\lambda\lambda$4959, 5007, and [\ion{S}{2}] $\lambda\lambda$6716, 6731. By fitting all these emission features with Gaussian profiles, we obtain an average redshift of $z=0.0307\pm0.0003$. With the diagnostic emission-line equivalent width (EW) ratios ($\log$~[\ion{N}{2}]/H$\alpha=-1.05\pm0.08$ and $\log$~[\ion{O}{3}]/H$\beta=0.19\pm0.02$),\footnote{Here [\ion{N}{2}] denotes the EW of the [\ion{N}{2}] $\lambda$6583 line, and [\ion{O}{3}] denotes the EW of the [\ion{O}{3}] $\lambda$5007 line.} the host is consistent with star-forming galaxies in the \citet[][hereafter BPT]{BPT_1981} diagram \citep[see also][]{Veilleux_1987}. Additional discussion of the host galaxy's properties is presented in Section~\ref{sec:host}.

To estimate the distance modulus of \sn, we first use the 2M++ model \citep{Carrick2015_2M++} to estimate the peculiar velocity of its host galaxy, PSO\,J175312.663+005122.078, to be $179\pm250$\,\kms. This, combined with the recession velocity in the frame of the cosmic microwave background\footnote{\url{https://ned.ipac.caltech.edu/velocity_calculator}} (CMB) $v_\mathrm{CMB}=9136$\,\kms, yields a net Hubble recession velocity of $9307\pm250$\,\kms. Adopting $H_0=70$\,\kms\,Mpc$^{-1}$, $\Omega_M=0.3$, and $\Omega_\Lambda=0.7$, we estimate the luminosity distance of \sn\ to be 136.1\,Mpc, equivalent to a distance modulus of $\mu=35.67\pm0.06$\,mag.

To evaluate the potential host-galaxy extinction, we measure the Balmer decrement and find the flux ratio of H$\alpha$ to H$\beta$ to be $3.26\pm0.13$, while the theoretical, extinction-free value is 2.86 \citep[assuming case B recombination;][]{Osterbrock_2006}. Using the extinction law from \citet{Fitzpatrick1999} and assuming $R_V=3.1$, this yields $E(B-V)=0.11\pm0.04$\,mag. This result is consistent with a model of the host galaxy's spectral energy distribution (SED; illustrated in Section~\ref{sec:host}), $E(B-V)=0.13\pm0.01$\,mag. As we do not know the precise location of \sn\ within its host galaxy, we adopt these reddening values as an upper limit to the total host-galaxy reddening. 

\subsection{Optical Photometry}
\sn\ was monitored in the $g_\mathrm{ZTF}$ and $r_\mathrm{ZTF}$ bands by ZTF as part of its ongoing Northern Sky Survey \citep{Bellm_ZTF_2019b}. We adopt a Galactic extinction of $E(B-V)_\mathrm{MW}=0.404\,$mag \citep{Schlafly2011}, and correct all photometry using the \citet{Fitzpatrick1999} extinction model. The host extinction is not well constrained. While the potential host extinction could be up to $E(B-V)_\mathrm{host}\approx0.13$, the lack of \ion{Na}{1}\,days absorption at the redshift of the host galaxy is consistent with no additional host extinction, though see \citet{Poznanski_2011} for caveats on the use of \ion{Na}{1}\,days absorption as a proxy for extinction. Thus throughout the paper, we adopt a fiducial assumption of no host extinction and discuss the possible effects of addition extinction on constraining the progenitor properties in Section~\ref{sec:model}. Unless otherwise specified, the data displayed in the figures are only corrected for Galactic extinction.

\begin{figure*}
    \centering
    \includegraphics[width=\textwidth]{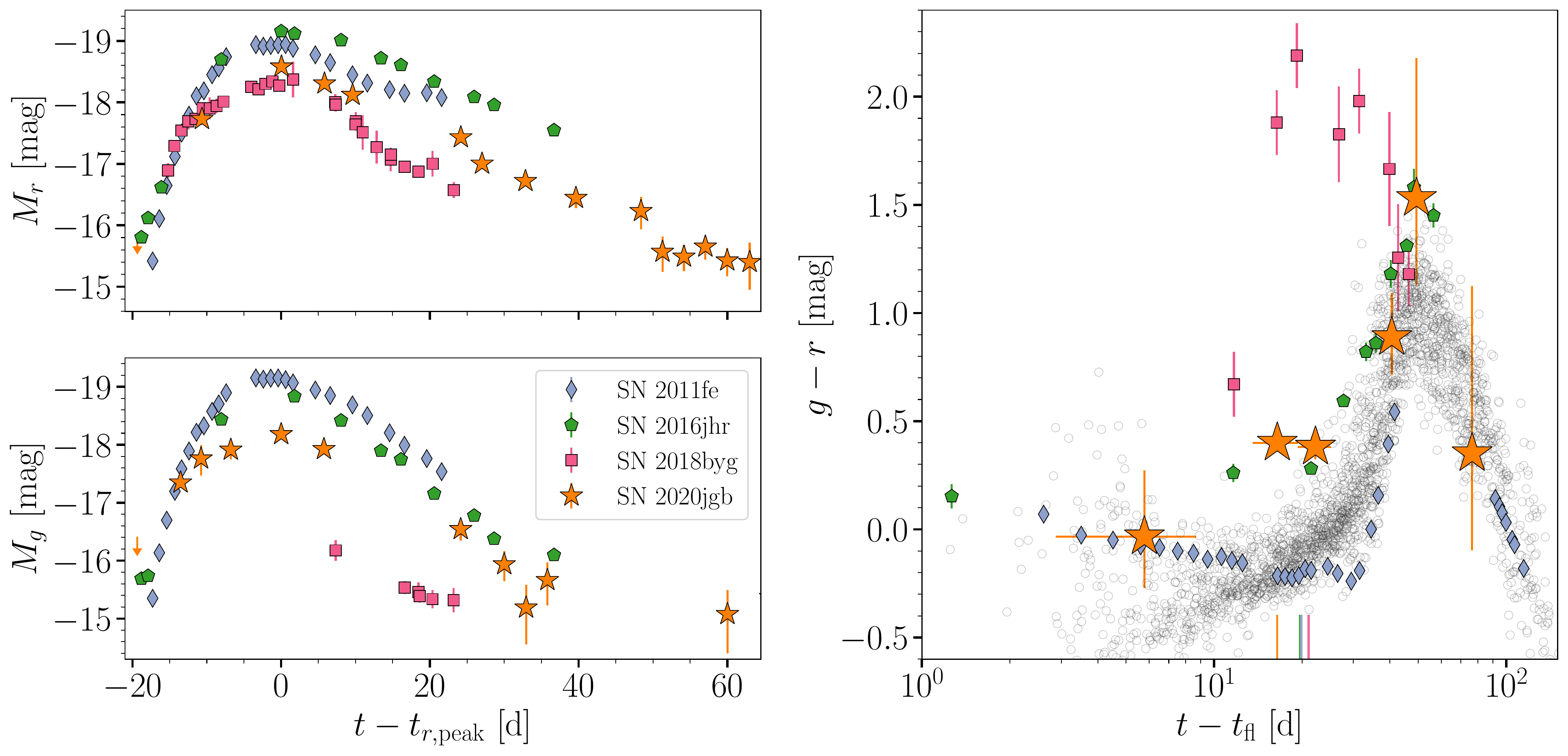}
    \caption{Comparison of the photometric properties of \sn\ with those of SN\,2011fe \citep[normal SN\,Ia;][]{Pereira_2013}, SN\,2016jhr \citep[normal-luminosity He-shell DDet;][]{jiang_16jhr_2017}, and SN\,2018byg \citep[subluminous He-shell DDet;][]{de_18byg_2019}. \textit{Left}: multiband light curves. The upper (lower) panel shows the evolution in the $r$-band ($g$-band) absolute magnitude. {The arrows mark the 5$\sigma$ limit of the last nondetections of \sn\ in $g_\mathrm{ZTF}$ and $r_\mathrm{ZTF}$.} \textit{Right}: $g-r$ color evolution. For each object, the peak epoch is marked by a vertical line with the corresponding color on the bottom axis. The gray circles denote the $g_\mathrm{ZTF}-r_\mathrm{ZTF}$ color evolution of 62 normal SNe\,Ia (open circles) with prompt observations within 5\,days of first light by ZTF \citep{Bulla2020}. }
    \label{fig:photometry}
\end{figure*}

The forced-photometry absolute light curves\footnote{\url{https://web.ipac.caltech.edu/staff/fmasci/ztf/forcedphot.pdf}} in $g_\mathrm{ZTF}$ and $r_\mathrm{ZTF}$ are shown in Figure~\ref{fig:photometry}, where we display all measurements having a signal-to-noise ratio (S/N) greater than 2. The light curves are reduced using the pipeline from Miller et al. (2022, in preparation); see also \citet{Yao_2019}.

\subsection{Optical Spectroscopy}\label{sec:optical_spec}

\begin{figure*}
    \centering
    \includegraphics[width=\textwidth]{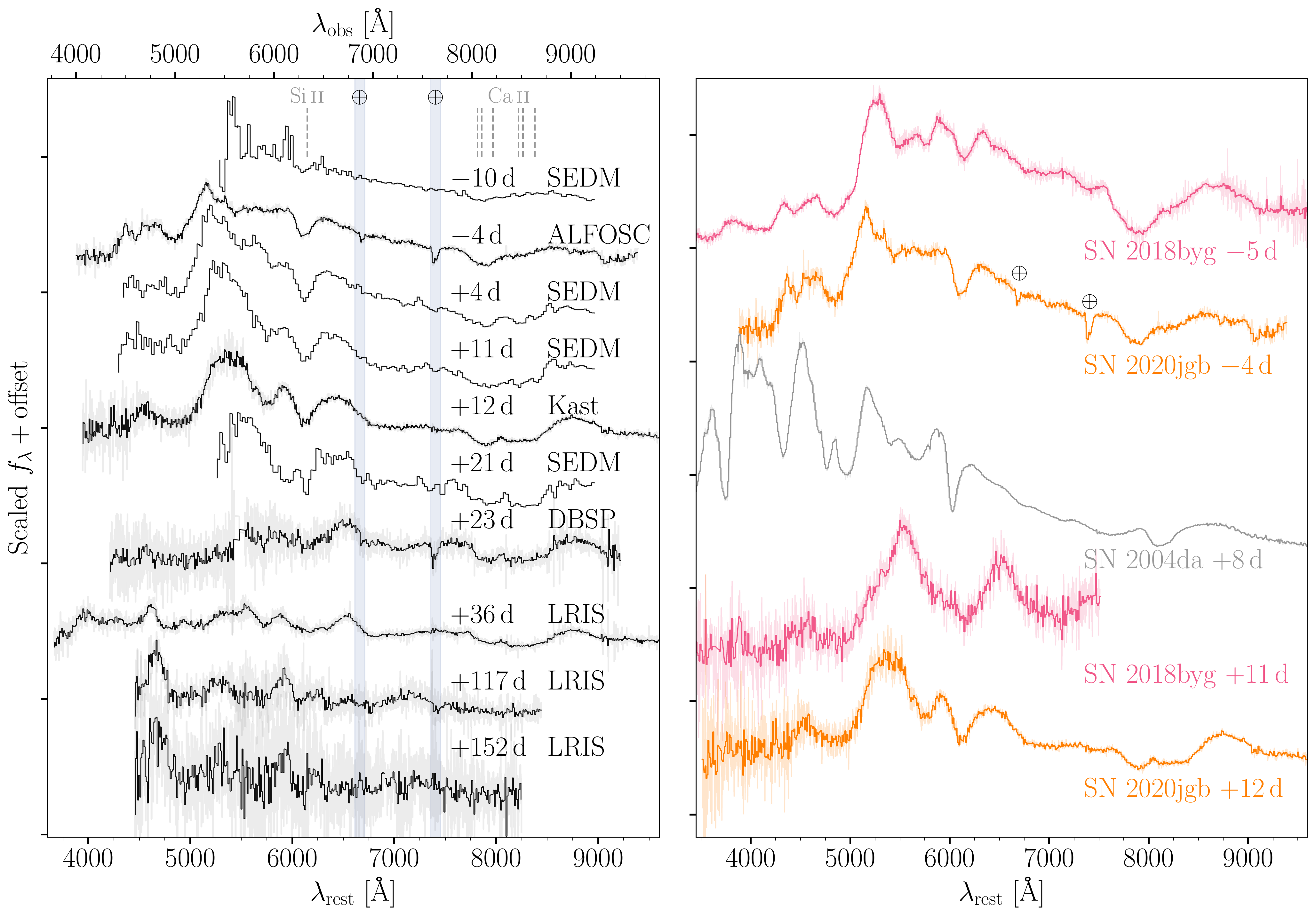}
    \caption{The optical spectral evolution of \sn\ is typical of that of a peculiar SN\,Ia triggered by an He-shell DDet. \textit{Left}: optical spectral sequence of \sn. Rest-frame phases (days) relative to the $r_\mathrm{ZTF}$-band peak and instruments used are posted next to each spectrum. Spectra have been corrected for $E(B-V)_\mathrm{MW} = 0.404$\,mag and are shown in gray. The black lines are binned spectra with a bin size of 10\,\AA, except for the SEDM spectra, whose resolution is lower than the bin size. In the last two spectra, we have subtracted the light from the host galaxy. Only regions with $\mathrm{S/N}>2.5$ after binning are plotted. The corresponding wavelengths of the \ion{Si}{2} $\lambda$6355 line (with an expansion velocity of 10,000\,\kms) and the \ion{Ca}{2} IRT (with expansion velocities of both 10,000\,\kms and 25,000\,\kms) are marked by the vertical dashed lines.
    \textit{Right}: spectral comparison with SN\,2018byg \citep[subluminous He-shell DDet;][]{de_18byg_2019} and SN\,2004da \citep[normal luminosity;][]{Silverman_2012}.}
    \label{fig:spec_evo}
\end{figure*}

We obtained optical spectra of the object from $\sim$$-10$\,days to $\sim$+150\,days relative to the $r_\mathrm{ZTF}$-band peak\footnote{{Unless otherwise specified, the ``peak'' or ``maximum brightness'' of \sn\ refers to its maximum $r_\mathrm{ZTF}$-band brightness.}}, using the Spectral Energy Distribution Machine \citep[SEDM;][]{SEDM_2018} on the automated 60\,inch telescope \citep[P60;][]{P60_2006} at Palomar Observatory, the Kast Double Spectrograph \citep{miller1994kast} on the Shane 3\,m telescope at Lick Observatory, the Andalucia Faint Object Spectrograph and Camera (ALFOSC)\footnote{\url{https://www.not.iac.es/instruments/alfosc/}} installed at the Nordic Optical Telescope (NOT), the Double Beam Spectrograph (DBSP) on the 200\,inch Hale telescope \citep[P200;][]{P200_1982}, and the Low Resolution Imaging Spectrometer (LRIS) on the Keck-I 10\,m telescope \citep{Keck_1995}. With the exception of observations obtained with SEDM, all spectra were reduced using standard procedures \citep[e.g.,][]{Matheson_2000}. The SEDM spectra were reduced using the custom \texttt{pysedm} software package \citep{Rigault_pysedm_2019}. Details of the spectroscopic observations are listed in Table~\ref{tab:spec}, and the resulting spectral sequence is shown in Figure~\ref{fig:spec_evo}. All the spectra listed in Table~\ref{tab:spec} will be available on WISeREP \citep{wiserep_2012}.

\begin{deluxetable}{lrcccc}
\tabletypesize{\scriptsize}
\tablewidth{0pt}
\tablecaption{Spectroscopic observations of \sn\label{tab:spectra} and the host galaxy.}
\tablehead{
\colhead{$t_\mathrm{obs}$} &
\colhead{Phase} &
\colhead{Telescope/} &
\colhead{$R$} &
\colhead{Range} &
\colhead{Airmass} \\
\colhead{(MJD)} &
\colhead{(days)} &
\colhead{Instrument} &
\colhead{$(\lambda/\Delta\lambda)$} &
\colhead{(\AA)} & 
\colhead{}
}
\startdata
58,976.42 &  $-$9.7 & P60/SEDM & 100 & 3770--9220 & 1.23\\
58,982.12 & $-$4.2 & NOT/ALFOSC & 360 & 4000--9620 & 1.17\\
58,990.43 &  $+$3.9 & P60/SEDM & 100 & 3770--9220 &  1.23\\
58,997.44 & $+$10.7 & P60/SEDM & 100 & 3770--9220 &  1.29\\
58,998.41 & $+$11.6 & Shane/Kast & 750 & 3620--10720 & 1.28\\ 
59,008.41 & $+$21.3 & P60/SEDM & 100 & 3770--9220 & 1.28\\
59,009.45 & $+$22.4 & Gemini-N/GNIRS & 1800 & 8230--25150 &1.07\\
59,010.40 & $+$23.3 & P200/DBSP & 700 & 3200--9500 &  1.27\\
59,023.58 & $+$36.1 & Keck I/LRIS & 1100 & 3200--10250 & 2.04\\
59,107.29 & $+$117.3 & Keck I/LRIS & 1100 & 3200--10250 & 1.31\\
59,143.26 & $+$152.2 & Keck I/LRIS & 1100 & 3200--10250 & 2.16\\
59,669.60 & host & Keck II/DEIMOS & 2100 & 4500--8700 & 1.14\\ 
\enddata
\tablecomments{Phase is measured relative to the $r_\mathrm{ZTF}$-band peak in the rest frame of the host galaxy. The resolution $R$ is reported for the central region of the spectrum.}
\label{tab:spec}
\end{deluxetable}

\subsection{Near-Infrared Spectroscopy}
We obtained one NIR (0.8--2.5\,\micron) spectrum of \sn\ using the Gemini near-infrared spectrometer \citep[GNIRS;][]{GNIRS1998} on the Gemini North telescope on 2020 June 9 ($\sim$22\,days after $r_\mathrm{ZTF}$-band peak), with a total integration time of 2400\,s. The GNIRS spectrum was reduced with \texttt{PypeIt}.

\begin{figure*}
    \centering
    \includegraphics[width=\textwidth]{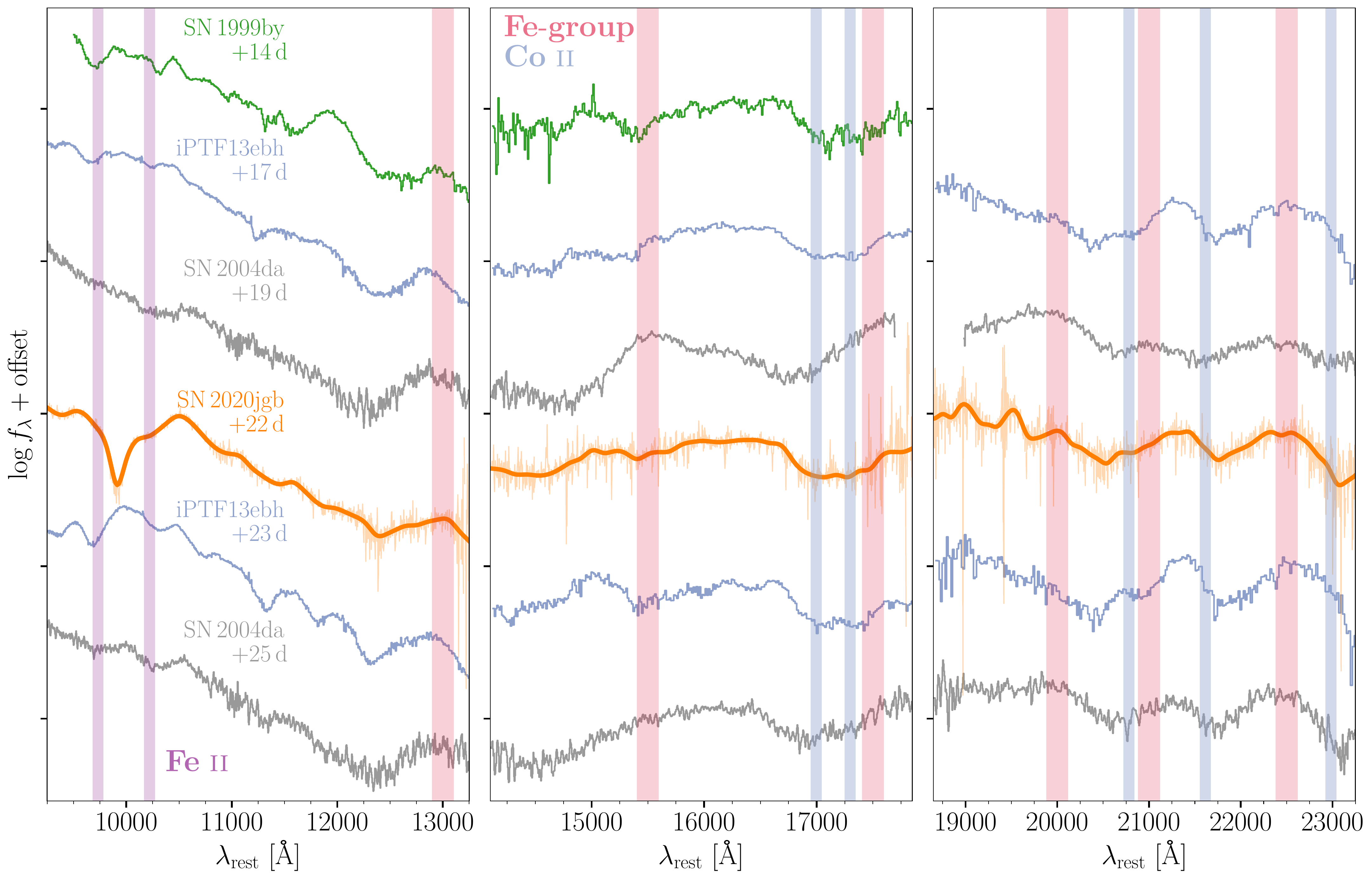}
    \caption{NIR spectra of \sn, {a normal-luminosity SNe\,Ia \citep[SN\,2004da;][]{Marion2009_NIR}, and two subluminous SNe\,Ia (SN\,1999by and iPTF13ebh; \citealp{Hoeflich_2002}; \citealp{Hsiao_13ebh_2015}), obtained 15--25\,days after maximum brightness. All the objects show} similar spectral features except the absorption line near 1\,\micron. For each spectrum, the continuum at $\gtrsim$1.2\,\micron\ is significantly reshaped by the line-blanketing from Fe-group elements (red stripes), which are continuous emission features composed of unresolved Fe-group lines peaking at $\sim$1.30, 1.55, 1.75, 2.00, 2.10, and 2.25\,\micron\ \citep{Marion2009_NIR}. Between these peaks lie multiple strong \ion{Co}{2} absorption lines (blue stripes), for which a typical post-maximum expansion velocity of 8000\,\kms\ is assumed. The purple stripes correspond to \ion{Fe}{2} $\lambda$9998 and \ion{Fe}{2} $\lambda$10500, also with an expansion velocity of 8000\,\kms.}
    \label{fig:NIR_spec}
\end{figure*}

\section{Analysis} \label{sec:analysis}
\subsection{Photometric Properties} \label{sec:phot_analysis}
\sn\ exhibited a fainter light curve than normal SNe\,Ia. In Figure~\ref{fig:photometry}, we compare the photometric properties of \sn\ with the nearby, well-observed SN\,2011fe in $g_\mathrm{ZTF}$ and $r_\mathrm{ZTF}$ synthetic photometry from the spectrophotometric time series of \citet{Pereira_2013}, as well as two {peculiar He-shell DDet events}, including the normal-luminosity event SN\,2016jhr \citep{jiang_16jhr_2017} and the subluminous event SN\,2018byg \citep{de_18byg_2019}. All of these light curves have been corrected for Galactic reddening, while $K$-corrections have not been performed,\footnote{These SNe were all observed in slightly different $g$ and $r$ filters.} because we do not have complete spectral sequences of these peculiar events.

While the observational coverage is sparse in the rise to maximum light, from Figure~\ref{fig:photometry} it is clear that \sn\ is less luminous than normal SNe\,Ia (e.g., SN\,2011fe). If the host galaxy reddens \sn\ by $E(B-V)_\mathrm{host} = 0.13$\,mag, then \sn\ would be $\sim$0.3\,mag brighter in the $r_\mathrm{ZTF}$ band and $\sim$0.5\,mag in the $g_\mathrm{ZTF}$ band, making it comparable to SN\,2011fe in $r_\mathrm{ZTF}$, yet still $\sim$0.5\,mag fainter in $g_\mathrm{ZTF}$.
In the right panel of Figure~\ref{fig:photometry}, we compare the color evolution ($g-r$) of these objects relative to the measured time of first light \tfl, accompanied by 62 normal SNe\,Ia (open circles) observed within 5 days of \tfl\ by ZTF \citep[from][]{Bulla2020}. They have been corrected for Galactic extinction, but $K$-corrections have not been performed for consistency. For \sn, the early rise of the light curve was not well sampled, so we estimate \tfl\ as the midpoint of the first detection and the last nondetection. We adopt an uncertainty in this estimate of 3\,days. All three He-shell DDet candidates are undoubtedly redder than normal SNe\,Ia. At maximum light, \sn\ ($g_\mathrm{ZTF}-r_\mathrm{ZTF}\approx0.4$\,mag) was not as red as SN\,2018byg ($g-r\approx2.2$\,mag), but exhibited a similar color as SN\,2016jhr ($g-r\approx0.3$\,mag). Adopting $E(B-V)_\mathrm{host} = 0.13$\,mag still results in a relatively red color for \sn\ ($g_\mathrm{ZTF}-r_\mathrm{ZTF}\approx0.2$\,mag) compared to normal SNe\,Ia ($g_\mathrm{ZTF}-r_\mathrm{ZTF}\approx-0.1$\,mag).

Interestingly, for both SN\,2018byg and \sn, near their maximum light the spectra sharply peak at $\sim$5200\,\AA\ in the SN rest frame (see Figure~\ref{fig:spec_evo}), which is close to the red edge of the $g$/$g_\mathrm{ZTF}$ filter ($\sim$4000--5500\,\AA). Thus, modest redshifts ($z\gtrsim0.03$) can produce significant $K$-corrections, which constitute a substantial fraction of the observed red $g-r$ colors for these events. For \sn, using the ALFOSC spectrum obtained at $-$4\,days, we estimate the $K$-correction to be $K_{g-r}\approx -0.2$\,mag, the $g-r$ color being bluer in the rest frame. SN\,2018byg is at a higher redshift ($z=0.066$) so the $K$-correction is more extreme ($K_{g-r}\approx-1.0$\,mag). Future efforts to identify additional subluminous He-shell DDet candidates can utilize the red $g-r$ color to improve their search efficiency.

\subsection{Optical Spectral Properties}
In Figure~\ref{fig:spec_evo}, we show the optical spectral sequence of \sn, and compare its spectra with those of some other SNe\,Ia at similar phases relative to peak brightness. For the spectra obtained after $+100$\,days there is clear contamination from the host galaxy, including the presence of narrow emission lines. For these spectra we subtract the galaxy light as measured in the DEIMOS spectrum from 2022 (see Section~\ref{sec:optical_spec}). The earliest spectrum was obtained by SEDM $\sim$10\,days before $r_\mathrm{ZTF}$-band peak. We only show portions of the binned spectrum where the $\mathrm{S/N}>2.5$. The continuum is almost featureless with some marginal detection of \ion{Si}{2} $\lambda$6355 at $\sim$6100\,\AA, the hallmark of SNe\,Ia. {The subsequent spectra show a strong suppression of flux blueward of $\sim$5000\,\AA,\footnote{This feature is prominent in \sn\ when we adopt $E(B-V)_\mathrm{host} = 0.0$ or $0.13$\,mag.} one of the major differences from those of normal SNe\,Ia. The} \ion{Si}{2} features become more prominent and are clearly detected until $\sim$12\,days after maximum light. We measure \ion{Si}{2} expansion velocities following a procedure similar to that of  \citet{Childress_2013,Childress_2014} and \citet{Maguire_2014}. The fitting region is selected by visual inspection. The continuum is assumed to be linear, and the absorption profile after the continuum normalization is assumed to be composed of double Gaussian profiles centered at 6347\,\AA\ and 6371\,\AA. Within the model, the continuum flux densities at the blue and red edges are free parameters for which we adopt a normal distribution as a prior. The mean and standard deviation for the distribution are the observed flux density and its uncertainty (respectively) at each edge of the fitting region. Three more parameters (amplitude, mean velocity, logarithmic velocity dispersion) are used to characterize the double Gaussian profile, whose priors are set to be flat. This means the depths and widths of both peaks are forced to be the same, as \citet{Maguire_2014} adopted in the optically thick regime. The posteriors of the five parameters are sampled simultaneously with \texttt{emcee} \citep{emcee_2013} using the Markov Chain Monte Carlo method. We find that the mean expansion velocity is $\sim$11,500\,\kms\ near maximum light.

In many SNe\,Ia, the \ion{Ca}{2} near-infrared triplet (\ion{Ca}{2} IRT) $\lambda\lambda$8498, 8542, 8662 causes two distinct components \citep{Mazzali_2005}, which are conventionally referred to as photospheric-velocity features (PVFs) and high-velocity features (HVFs). The PVFs originate from the main line-forming region with typical photospheric (i.e., bulk ejecta) velocities, while the HVFs are blueshifted to much shorter wavelengths, indicating significantly higher (by $\gtrsim$6000\,\kms) velocities than typical PVFs \citep{Silverman_HVF_2015}. Figure~\ref{fig:spec_evo} shows that \sn\ has prominent HVFs of \ion{Ca}{2} IRT. The HVFs are visible in our first spectrum of \sn\ at $-10$\,days, and remain prominent through $+36$\,days. Using the same technique we use to model the \ion{Si}{2} features, we fit the HVFs and PVFs simultaneously. Both are fit by multiple Gaussian profiles assuming each line in the triplet can be approximated by the same profile (i.e., same amplitude and velocity dispersion). A best-fit expansion velocity of HVFs at $-10$\,days is $\sim$26,000\,\kms. In the spectrum at $-4$\,days, we observe a clear delineation between the HVFs and PVFs. For this and subsequent spectra, we fit the broad absorption features with two different velocity components simultaneously. From $-4$ to $+23$\,days, the speed of the HVFs declines slightly to $\sim$24,000\,\kms, and the speed of PVFs declines from $\sim$11,000\,\kms\ to $\sim$9,000\,\kms. As in normal SNe\,Ia, the relative strength between the HVFs and PVFs decreases with time. {In Table~\ref{tab:vel} we report the evolution of the expansion velocity of \ion{Si}{2} $\lambda$6355 and the \ion{Ca}{2} IRT.}
\begin{deluxetable}{rcccc}
    \tabletypesize{\scriptsize}
    \tablewidth{0pt}
    \tablecaption{{The evolution of the expansion velocity of \ion{Si}{2} $\lambda$6355 and \ion{Ca}{2} IRT\label{tab:vel} and the velocity of the 1\,\micron\ feature assuming a \ion{He}{1} origin in the GNIRS spectrum.}}
    \tablehead{
    \colhead{Phase} &
    \colhead{$v_{\mathrm{Si}\,\textsc{ii}}$} &
    \colhead{$v_{\mathrm{Ca}\,\textsc{ii},\mathrm{PVF}}$} &
    \colhead{$v_{\mathrm{Ca}\,\textsc{ii},\mathrm{HVF}}$} &
    \colhead{$v_{\mathrm{He}\,\textsc{i}?}$} \\
    \colhead{(d)} &
    \colhead{($10^3$\,\kms)} &
    \colhead{($10^3$\,\kms)} &
    \colhead{($10^3$\,\kms)} &
    \colhead{($10^3$\,\kms)}
    }
    \startdata
    $-$9.7 & $\ldots$          & $15.33\pm0.12$ & $25.88\pm0.25$ & $\ldots$\\
    $-$4.2 & $11.737\pm0.073$  & $11.18\pm0.23$ & $24.22\pm0.11$ & $\ldots$\\
    $+$3.9 & $11.356\pm0.085$  & $9.76\pm0.21$  & $24.17\pm0.30$ & $\ldots$\\
    $+$10.7 & $11.53\pm0.20$   & $8.8\pm1.0$    & $24.0\pm1.1$   & $\ldots$\\
    $+$11.6 & $11.239\pm0.075$ & $9.40\pm0.13$  & $24.58\pm0.11$ & $\ldots$\\
    $+$21.3 & $10.46\pm0.21$   & $9.10\pm0.59$  & $24.16\pm0.69$ & $\ldots$\\
    $+$22.4 & $\ldots$         & $\ldots$       & $\ldots$       & $26.178\pm0.062$\\
    $+$23.3 & $\ldots$         & $9.57\pm0.46$  & $23.59\pm0.47$ & $\ldots$\\
    $+$36.1 & $\ldots$         & $9.57\pm0.15$  & $23.84\pm0.15$ & $\ldots$\\
    \enddata
    \label{tab:spec}
    \end{deluxetable}

We obtained two LRIS spectra at $+117$\,days and $+152$\,days, both of which are dominated by Fe-group elements and resemble those of normal SNe\,Ia \citep[e.g., SN\,2011fe;][]{Mazzali_2015}, showing some enhancement in flux between $\sim$4500 and $\sim$6000\,\AA. There are no signs of emission due to the [\ion{Ca}{2}] $\lambda\lambda$7291, 7324 doublet.

\sn\ does not show any absorption features associated with \ion{O}{1} $\lambda$7774. While the low luminosity, red color, absence of hydrogen features, and star-forming host galaxy of \sn\ are also reminiscent of Type Ic SNe (SNe\,Ic), which arise from stripped-envelope massive stars, SNe\,Ic usually exhibit stronger \ion{O}{1} $\lambda$7774 lines. The ratio of the relative line depths\footnote{The relative depth is defined as the absorption line depth relative to the pseudo-continuum. See \citet{Sun_2017} for more details.} between the \ion{O}{1} $\lambda$7774 line and the \ion{Si}{2} $\lambda$6355 line is expected to be greater than 1 in typical SNe\,Ic \citep{Gal-Yam_2017, Sun_2017}. {\sn\ additionally does not show [\ion{O}{1}] or [\ion{Ca}{2}] emission lines in the nebular phase, which are ubiquitous in SNe\,Ic \citep{Jerkstrand_2017}}. Consequently, we can definitively conclude \sn\ is not an SN\,Ic.

{The observational properties of \sn\ are distinct from those of other subluminous thermonuclear SNe, including SN\,2002cx-like \citep[02cx-like\footnote{This subclass is also referred to as Type Iax SNe \citep{Foley_Iax_2013}};][]{Li_02cx_2003}, SN\,1991bg-like \citep[91bg-like;][]{Filippenko_91bg_1992}, and SN\,2002es-like \citep[02es-like;][]{Ganeshalingam_02es_2012} objects. The 02cx-like subclass is known to show a much bluer color near peak luminosity \citep[$g-r\approx0$\,mag;][]{Miller_2017} than \sn. While 91bg-like and 02es-like SNe are redder than normal SNe\,Ia (due to the \ion{Ti}{2} absorption trough at $\sim$4200\,\AA), they exhibit significant emission blueward of $\sim$5000\,\AA. They also do not exhibit HVFs of \ion{Ca}{2} IRT \citep[e.g.,][]{Silverman_HVF_2015}, in contrast to \sn.}

The optical spectral evolution of \sn\ resembles that of SN\,2018byg, a subluminous He-shell DDet SN. At early times, both SNe were relatively blue and featureless, with broad and shallow \ion{Ca}{2} IRT absorption. As they evolved closer to maximum light, they developed strong continuous absorption blueward of $\sim$5000\,\AA. Meanwhile, \ion{Si}{2} $\lambda$6355 and the \ion{Ca}{2} IRT became more prominent. Neither \ion{O}{1} nor \ion{S}{2} was detected in either object. In the He-shell DDet scenario, a large amount of Fe-group elements would be synthesized in the shell, which would cause significant line-blanketing near maximum light \citep{Kromer_DD_2010, polin_observational_2019} and high-velocity intermediate-mass elements like \ion{Ca}{2} \citep{Fink_DD_2010, Kromer_DD_2010,Shen_DD_2014}. The similarity to SN\,2018byg makes \sn\ another promising He-shell DDet SN candidate.

\subsection{NIR Spectral Properties}
\label{sec:NIR_spec}
The NIR spectrum of \sn\ is compared with {those of a normal SNe\,Ia (SN\,2004da; data from \citealp{Marion2009_NIR}) and two subluminous SNe\,Ia (SN\,1999by and iPTF13ebh; \citealp{Hoeflich_2002} and \citealp{Hsiao_13ebh_2015}) at a similar phase in Figure~\ref{fig:NIR_spec}.} \sn\ shows a strong absorption feature at $\sim$0.99\,\micron, which is not seen in normal SNe\,Ia. This feature was still significant two weeks later, as detected with LRIS on Keck (see Figure~\ref{fig:hvf_comp}), though it was only partially covered. Aside from this prominent feature, \sn\ resembles normal SNe\,Ia in the NIR. The shape of the continuum redward of $\sim$1.2\,\micron\ is significantly altered by line-blanketing from Fe-group elements. Just like normal {and other subluminous} SNe\,Ia, \sn\ shows an enhancement of flux at about 1.30, 1.55, 2.00, 2.10, and 2.25\,\micron, accompanied by several \ion{Co}{2} absorption lines. It is especially similar to SN\,2004da at +25\,days as the steep increase in flux at $\sim$1.55\,\micron, known as the \textit{H}-band break \citep{Hsiao_CSP_2019}, has become less prominent. To summarize, the NIR spectrum of \sn\ is dominated by Fe-group elements, consistent with the nucleosynthetic yield of a WD thermonuclear explosion. However, the  1\,\micron\ feature adds to the peculiarities of \sn\ as an SN\,Ia.

\citet{Marion2009_NIR} presented a sample of 15 NIR spectra of normal SNe\,Ia between +14 and +75\,days relative to maximum light, and none of those spectra show prominent absorption features around 1\,\micron. We have investigated several potential identifications for this feature (see below), none of which provides a completely satisfying explanation.

The most tantalizing possibility is that the absorption is due to \ion{He}{1} $\lambda$10830. Modern DDet models reveal that part of the helium in the shell will be left unburnt \citep[e.g.,][]{Kromer_DD_2010,Woosley_2011,polin_observational_2019}. With full non-local-thermodynamic-equilibrium (nLTE) physics taken into consideration, \ion{He}{1} features are unambiguously expected in some He-shell DDet SNe, among which \ion{He}{1} $\lambda$10830 is the most prominent absorption line \citep{Dessart_2015,Boyle2017_Helium}.\footnote{Since helium has high excitation states, optical and NIR helium lines require nonthermal excitation \citep[e.g., collision with fast electrons;][]{Lucy_1991}. Models assuming LTE radiative transfer neglect nonthermal effects; thus, they are not able to characterize the helium features.} 
Figure~\ref{fig:hvf_comp} shows that the 1\,\micron\ feature, if associated with \ion{He}{1} $\lambda$10830, has a velocity of $\sim$26,000\,\kms {, which stays roughly the same from $\sim$22 to $\sim$36\,days after maximum light}. This speed is consistent with the unburnt helium in He-shell DDet models when the ejecta have reached homologous expansion \citep{Kromer_DD_2010, polin_observational_2019}, yet it is unclear whether the high-velocity unburnt helium could stay optically thick several weeks after maximum light. The \ion{Ca}{2} IRT also exhibits similarly high velocities at the same phase ($\sim$24,000\,\kms), meaning that high-velocity absorption is not impossible at this phase. The expansion velocity in the ejecta is roughly linearly proportional to the radius, so such a high velocity indicates that both the \ion{Ca}{2} IRT and the tentative \ion{He}{1} absorption line form far outside the normal photosphere, which has a velocity of only $\sim$10,000\,\kms. The two-dimensional (2D) models of \citet{Kromer_DD_2010} also suggest that helium may expand faster than the synthesized calcium in the He-shell. In this sense, the He-shell DDet scenario is supported because any unburnt helium would be located in the outermost ejecta.

We cannot claim an unambiguous detection of \ion{He}{1}, however, as our spectra lack definitive absorption from other \ion{He}{1} features that we would expect to be prominent, such as \ion{He}{1} $\lambda$20581. Considering a line velocity of $\sim$26,000\,\kms\ and a host-galaxy redshift of 0.0307, this line will be blueshifted to $\sim$1.95\,\micron\ in the observer frame, which overlaps with some strong telluric lines within 1.8--2.0\,\micron. In this region our NIR spectrum has S/N$\,\approx 5$ following telluric correction, yet we do not see any significant absorption feature. An upper limit of the equivalent width is determined to be $<$2\% that of the \ion{He}{1} $\lambda$10830 line, while the $\lambda$20581 line is theoretically supposed to be only a factor of 6--12 weaker, depending on the temperature \citep{Marion2009_NIR}. The observed 1\,\micron\ feature in \sn\ is as strong as the \ion{He}{1} $\lambda$10830 line in many helium-rich Type Ib supernovae \citep[SNe\,Ib;]{shahbandeh_carnegie_2021}. In SNe\,Ib, the \ion{He}{1} $\lambda$20581 line is weaker than the \ion{He}{1} $\lambda$10830 line, yet still prominent \citep{shahbandeh_carnegie_2021}. In one of the models of \citet{Boyle2017_Helium}, there is no obvious \ion{He}{1} $\lambda$20581 absorption in the synthetic spectra (see their Figure~7), but the model is intended to be representative of normal-luminosity SNe\,Ia. If the 1\,\micron\ feature is associated with \ion{He}{1}, it is unusual that we do not detect a corresponding feature around 2\,\micron.

Other possible identifications for the 1\,\micron\ feature include \ion{Mg}{2} $\lambda$10927, \ion{C}{1} $\lambda$10693, and \ion{Fe}{2} $\lambda$10500 and $\lambda$10863. The \ion{Mg}{2} $\lambda$10927 line is prevalent in the NIR spectra of SNe\,Ia, but usually disappears within a week after peak brightness \citep{Marion2009_NIR}. In \sn\ the 1\,\micron\ feature was still visible more than a month after maximum light in the Keck/LRIS spectrum. An \ion{Mg}{2} $\lambda$10927 identification would require an absorption velocity of $\sim$28,000\,\kms, $\sim$20\% faster than the HVFs of the \ion{Ca}{2} IRT at the same phase. Such a high-velocity \ion{Mg}{2} line has never been seen in other SNe\,Ia, and requires a high magnesium abundance in the outermost ejecta. However, the amount of magnesium synthesized in the detonation of the He-shell is expected to be tiny \citep{Fink_DD_2010,Kromer_DD_2010,polin_observational_2019,polin_nebular_2021}. On the other hand, if we attribute this 1\,\micron\ feature to high-velocity \ion{Mg}{2}, we would expect an even stronger \ion{Mg}{2} $\lambda$9227 line to be blueshifted to the red edge of the \ion{Ca}{2} IRT, which is not detected. Given the strength of the 1\,\micron\ feature, the \ion{Mg}{2} $\lambda$9227 line should not be completely obscured by the \ion{Ca}{2} IRT features.

\ion{C}{1} $\lambda$10693 is not observed as frequently as \ion{Mg}{2} $\lambda$10927 in SNe\,Ia. \citet{Hsiao_CSP_2019} presented a sample of five SNe\,Ia with \ion{C}{1} detections, showing that the \ion{C}{1} feature is strongest for fainter, fast-declining objects. However, in their sample, the \ion{C}{1} line is a pre-maximum feature which fades away as the luminosity peaks, so the discrepancy in phase is large. The required expansion velocity $\sim$22,000\,\kms\ is substantially faster than the estimated carbon velocity for the sample of \citet{Hsiao_CSP_2019} ($\sim$10,000--12,000\,\kms), but still consistent with the HVFs of the \ion{Ca}{2} IRT in \sn. Nonetheless, no significant carbon absorption is detected in the optical. It is also noteworthy that the amount of unburnt carbon is expected to be minimal in sub-\Mch WDs ignited by He-shell detonation \citep{polin_observational_2019}, in contrast to near-\Mch\ WDs ignited by pure deflagration where the carbon burning could be incomplete. We therefore would not expect to detect any carbon features in an He-shell DDet SN.

The \ion{Fe}{2} features in SNe\,Ia usually start to develop approximately three weeks after peak brightness, which is about the same phase as that at which we obtained our GNIRS spectrum. Two \ion{Fe}{2} lines, $\lambda$9998 and $\lambda$10500, are actually visible on the blue/red wings of the 1\,\micron\ feature (see Figure~\ref{fig:NIR_spec}). The \ion{Fe}{2} $\lambda$10863 line is not detected in the GNIRS spectrum. SN\,2004da shows very similar \ion{Fe}{2} features near 1\,\micron, in which \ion{Fe}{2} $\lambda$10500 is the strongest line at this phase, as displayed in Figure~\ref{fig:NIR_spec}. They correspond to an expansion velocity of $\sim$8000\,\kms, which is consistent with the PVFs of the \ion{Ca}{2} IRT at the same epoch. They also match the same two lines for normal SNe\,Ia \citep{Marion2009_NIR}, making the identification more reliable. Obviously, these two \ion{Fe}{2} features are wider and shallower than the strong feature between them. We fit the 1\,\micron\ feature with three Gaussian profiles. Two of them are set to be the blueshifted \ion{Fe}{2} $\lambda$9998 and $\lambda$10500, and the other is an uncorrelated Gaussian profile that mainly describes the deep absorption feature in the center of the line complex. We find that the shallower and wider \ion{Fe}{2} lines only make up $\sim$40\% of the total equivalent width, and the remaining $\sim$60\% comes from the central feature, which cannot be accounted for by any \ion{Fe}{2} feature at the same velocity. Given the similarity of the Fe-group line-blanketing between the GNIRS spectrum and the spectrum of SN\,2004da at +25\,days, the distribution of Fe-group elements inside each SN ejecta should be somewhat similar, so the central region of the 1\,\micron\ feature is not likely to be associated with \ion{Fe}{2} either.

\section{Discussion} \label{sec:discussion}
\subsection{Models} \label{sec:model}
\begin{figure*}
    \centering
    \includegraphics[width=\textwidth]{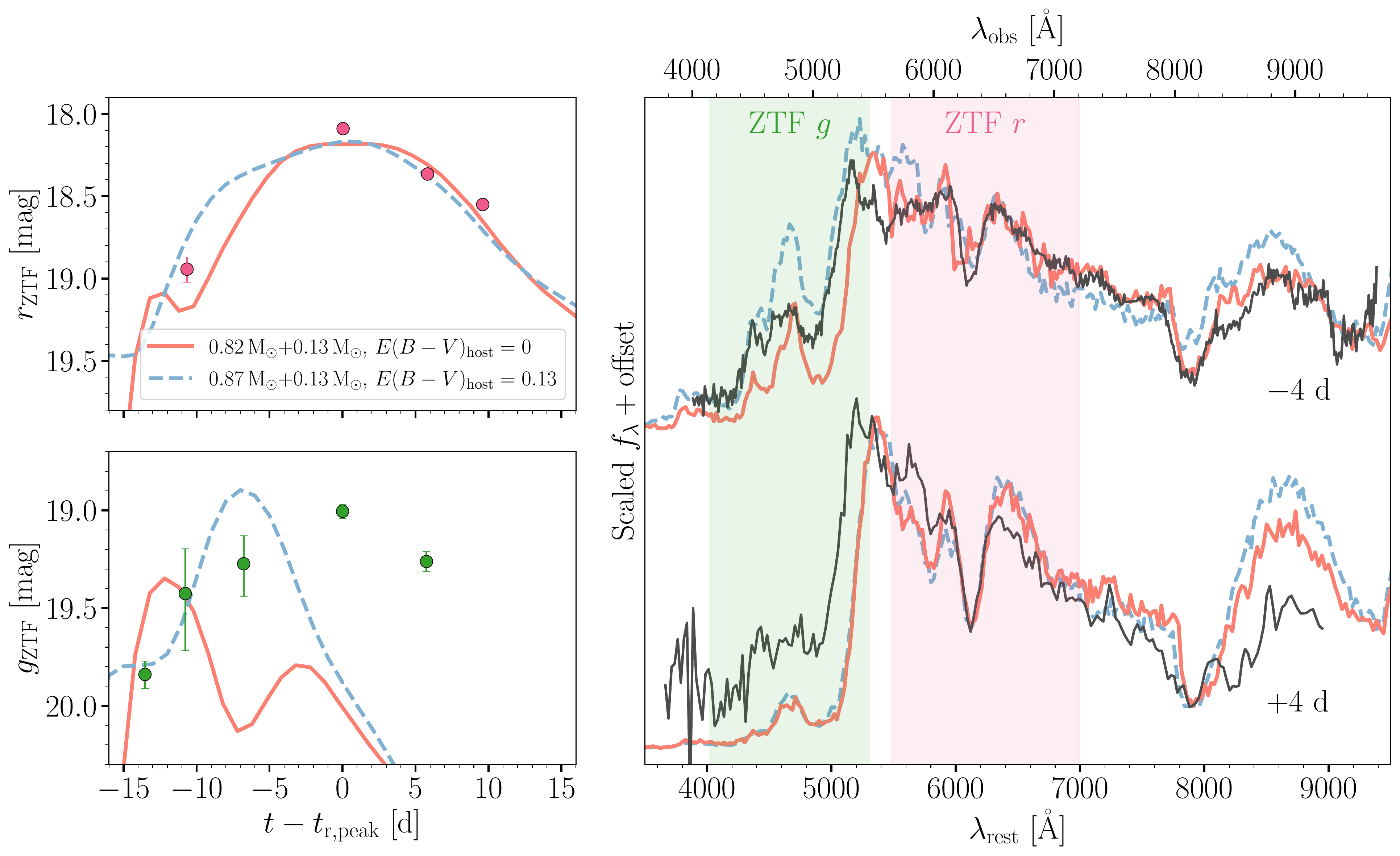}
    \caption{Spectrophotometric comparison of \sn\ observations with two He-shell DDet models using the methods described in \citet{polin_observational_2019}. For the $0.82\,\Msun+0.13\,\Msun$ model, only the Galactic reddening of $E(B-V)_\mathrm{MW}=0.404$\,mag is applied to the synthetic spectra and photometry; for the $0.87\,\Msun+0.13\,\Msun$ model, additional reddening of $E(B-V)_\mathrm{host}=0.13$\,mag from the host galaxy is assumed. {\it Left:} comparison of the ZTF photometry with the synthetic light curves. The model parameters are indicated in the legend as (C/O core mass $+$ He-shell mass). The upper (lower) panel shows the evolution in $r_\mathrm{ZTF}$ ($g_\mathrm{ZTF}$). The phases have been rescaled to the rest frame of the host galaxy. {\it Right:} comparison of the observed spectra with the models around maximum brightness. The shaded regions correspond to the coverage of the ZTF $g$ and $r$ filters with transmission above half-maximum. All spectra are normalized such that they would have the same synthetic brightness in $r_\mathrm{ZTF}$. The synthetic spectra are further binned with a size of 20\,\AA.}
    \label{fig:model}
\end{figure*}

We model \sn\ using the methods outlined by \citet{polin_observational_2019}; the process is twofold. After choosing an initial model that describes a WD of a given mass with a choice of He-shell mass, we use the \texttt{CASTRO} code \citep{Almgren_Castro_2010} to perform a 1D hydrodynamic simulation with simultaneous nucleosynthesis from the time of He-shell ignition through the secondary detonation and until the ejecta have reached homologous expansion ($\sim$10\,s). At this point we take the ejecta profile (velocity, density, temperature, and composition) and use the Monte Carlo radiative transport code \texttt{SEDONA} \citep{Kasen_Sedona_2006} to calculate synthetic light curves and spectra of our model under the assumption of LTE. 

For He-shell DDet SNe, the peak luminosity in $r_\mathrm{ZTF}$ is a proxy for the amount of \Ni\ synthesized in the detonation, which reflects the total progenitor mass \citep[C/O core $+$ He-shell;][]{polin_observational_2019}. We find that models with a total mass of $0.95\,\Msun$ reproduce the $r_\mathrm{ZTF}$-band peak brightness well if there is no extinction from the host galaxy. If the host galaxy reddens \sn\ by $E(B-V)_\mathrm{host}=0.13$\,mag, then specific luminosity in $r_\mathrm{ZTF}$ would be $\sim$25\% higher, and the corresponding progenitor mass would be roughly $1.00\,\Msun$. The uncertainty in the extinction limits the precision with which the progenitor mass of \sn\ can be constrained.

Nonetheless, the major photometric and spectroscopic features of \sn\ are consistent with those of a DDet SN with a massive shell. In Figure~\ref{fig:model}, we show the comparison of the observations of \sn\ with He-shell DDet models with a shell mass of $0.13\,\Msun$ and total masses of $0.95\,\Msun$ and $1.00\,\Msun$, respectively. To compare the models with the observations, we apply the adopted host reddening [$E(B-V)_\mathrm{host} = 0.0$ or $0.13$\,mag] to the rest-frame model SN spectrum, then redshift the model spectrum by 0.0307, before applying Galactic reddening [$E(B-V)_\mathrm{MW}=0.404$\,mag]. Both models reproduce the overall evolution of \sn\ in $r_\mathrm{ZTF}$, but fail to provide a reasonable fit to the light curve in $g_\mathrm{ZTF}$. Specifically, the peak brightness in $g_\mathrm{ZTF}$ is overestimated in the $0.87\,\Msun+0.13\,\Msun$ model but underestimated in the $0.82\,\Msun+0.13\,\Msun$ model. The {overall} $g_\mathrm{ZTF}$-band light curves in both models evolve faster than our observations, and quickly become $\gtrsim$1\,mag fainter than the observed $g_\mathrm{ZTF}$ brightness at the same epoch.

The spectral comparison reveals more details. We find that both models, especially the $0.82\,\Msun+0.13\,\Msun$ one, provide {a reasonable} match to the ALFOSC spectrum obtained $\sim$4\,days prior to the peak at rest-frame wavelengths $\gtrsim$5500\,\AA. The same is true for the SEDM spectrum obtained $\sim$4\,days after maximum brightness, though both models overpredict flux excess in the \ion{Ca}{2} IRT P-Cygni profile. Meanwhile, both models provide a poor fit to the observation in bluer regions. Before maximum brightness {in the observation}, the $0.87\,\Msun+0.13\,\Msun$ model exhibits weaker Fe-group line-blanketing, thus showing a much higher total flux in $g_\mathrm{ZTF}$. The $0.82\,\Msun+0.13\,\Msun$ model provides a proper level of line-blanketing, but the continuous absorption in the synthetic spectrum terminates at a longer wavelength ($\sim$5400\,\AA, as opposed to $\sim$5200\,\AA\ in the $-4$\,days spectrum). As we have already mentioned in Section~\ref{sec:phot_analysis} when discussing $K$-corrections, the observed flux in $g_\mathrm{ZTF}$ is extremely sensitive to the red edge of the line-blanketing region, which, in the observer frame, is close to the edge of the filter. Figure~\ref{fig:model} shows that while $f_\lambda$ peaks in the $g_\mathrm{ZTF}$ filter near maximum light, in the $0.82\,\Msun+0.13\,\Msun$ model the synthetic $f_\lambda$ peaks in the gap between the $g_\mathrm{ZTF}$ and $r_\mathrm{ZTF}$ filters. The same is true for the $0.87\,\Msun+0.13\,\Msun$ model after the $r_\mathrm{ZTF}$-band peak. Interestingly, this mismatch is also seen when fitting similar DDet models to SN\,2018byg \citep[see Figure~6 in][]{de_18byg_2019} despite the convincing match to observations at longer wavelengths, suggesting this is one of the systematics in our models. By manually shifting the synthetic spectra at $-$4\,days in the $0.82\,\Msun+0.13\,\Msun$ model blueward by 200\,\AA, we find that the corresponding synthetic magnitude in $g_\mathrm{ZTF}$ immediately increases by $\sim$0.5\,mag. Given the sensitivity of brightness in $g_\mathrm{ZTF}$ on modeling the line-blanketing and the uncertainty in our models from \citet{polin_observational_2019}, we do not attempt to fit the $g_\mathrm{ZTF}$-band light curve of \sn\ even near maximum light.

The systematics in modeling the line-blanketing (and the flux in many similar $g$ bands) may be attributed to a variety of factors on handling the explosion and radiative transfer. First, our models assume LTE, which is not valid once the ejecta become optically thin. Typically the bulk ejecta of a sub-\Mch\ SN\,Ia remain optically thick for $\sim$30\,days after the explosion. But in modeling the $g_\mathrm{ZTF}$-band brightness, the LTE assumption is more challenging because the major opacity in $g_\mathrm{ZTF}$ comes from the Fe-group line-blanketing in the outermost ejecta, where the optical depth may evolve differently from that near the photosphere. Hence, the LTE condition may become inapplicable much earlier. Furthermore, our 1D He-shell model is not capable of capturing multidimensional effects in the explosion such as asymmetries. The viewing angle is known to have a significant influence on the observed light curves \citep{Kromer_DD_2010, Sim_2012, Gronow_2020, Shen_2D_2021}, especially in bluer bands where the line-blanketing depends sensitively on the distribution of He-shell ashes \citep{Shen_2D_2021}. In previous studies of other He-shell DDet objects, the $g$-band brightness is systematically underpredicted shortly after the peak, despite the fact that redder bands can be fit decently \citep[e.g.,][]{jiang_16jhr_2017,jacobson-galan_16hnk_2020}.

Another discrepancy occurs in the late-time spectra. It is argued in \citet{polin_nebular_2021} that as the total progenitor mass in the He-shell DDet decreases, the SN gets fainter and the major coolants in the nebular phase change smoothly from Fe-group elements to the [\ion{Ca}{2}] $\lambda\lambda$7291, 7324 doublet.
For a total progenitor mass $\lesssim$1.0\,$\Msun$, [\ion{Ca}{2}] emission features are expected to dominate Fe-group features in the nebular phase, clearly in contrast to what we see in \sn. {Since there is no evidence that the progenitor mass of \sn\ is strongly underestimated (e.g., due to substantial host extinction that has not been accounted for), the absence of [\ion{Ca}{2}] emission features suggests} the transition between the Fe-strong and Ca-strong regimes {may occur} for a lower progenitor mass than simulations have predicted.  
{As for other peculiar DDet events,} SN\,2016hnk is estimated to have an even lower progenitor mass \citep[$\sim$0.87\,$\Msun$;][]{jacobson-galan_16hnk_2020} and shows [\ion{Ca}{2}] lines indisputably, drawing a lower limit of the progenitor mass for this transition \citep[see][for discussion on the potential host-galaxy extinction on SN\,2016hnk]{galbany_16hnk_2019}. {The late-time spectrum of SN\,2019ofm also exhibits prominent [\ion{Ca}{2}] emission.}\footnote{{For this reason, SN\,2016hnk and SN\,2019ofm also fall into the category of the calcium-rich (Ca-rich) transients, which are well known for their conspicuous nebular [\ion{Ca}{2}] emission \citep{Filippenko_2003, Perets_2010, Kasliwal_2012}. The strong [\ion{Ca}{2}] emission in the nebular phase may be explained by an He-shell DDet explosion \citep{Dessart_2015,polin_observational_2019}. See \citet{de_Ca_rich_2020} for a detailed discussion.}} SN\,2018byg and SN\,2016dsg were not observed in the nebular phase, so it remains unknown whether they exhibit [\ion{Ca}{2}] emission, which would otherwise be expected to show up at $\sim$$+150$\,days.

Given the strong match in the $r_\mathrm{ZTF}$-band light curves and the near-peak spectra at wavelengths $\gtrsim$5500\,\AA\ between the observations of \sn\ and our He-shell DDet models following \citet{polin_observational_2019}, we conclude that \sn\ is consistent with a DDet event ignited by a massive He-shell. {Our 1D LTE models cannot characterize the Fe-group line-blanketing effects accurately, leading to large uncertainty in the shell mass.} Readers are referred to our Appendix~\ref{app1} where we show the comparison of our observations to models with a variety of shell masses, in which the thinner-shell models (shell masses $<$0.1\,$\Msun$) cannot reproduce the properties of \sn. Depending on the extinction in the host galaxy, the total mass of the progenitor should be $\sim$$0.95$--$1.00\,\Msun$. To constrain the progenitor masses of additional He-shell DDet SNe to a higher precision, one should thoroughly discuss any potential host extinction. Multidimensional simulations with more realistic radiative transfer setups are necessary to resolve the systematics in our current models.

\subsection{The 1\,\micron\ Feature} \label{sec:1um}
\begin{figure}
    \centering
    \includegraphics[width=\linewidth]{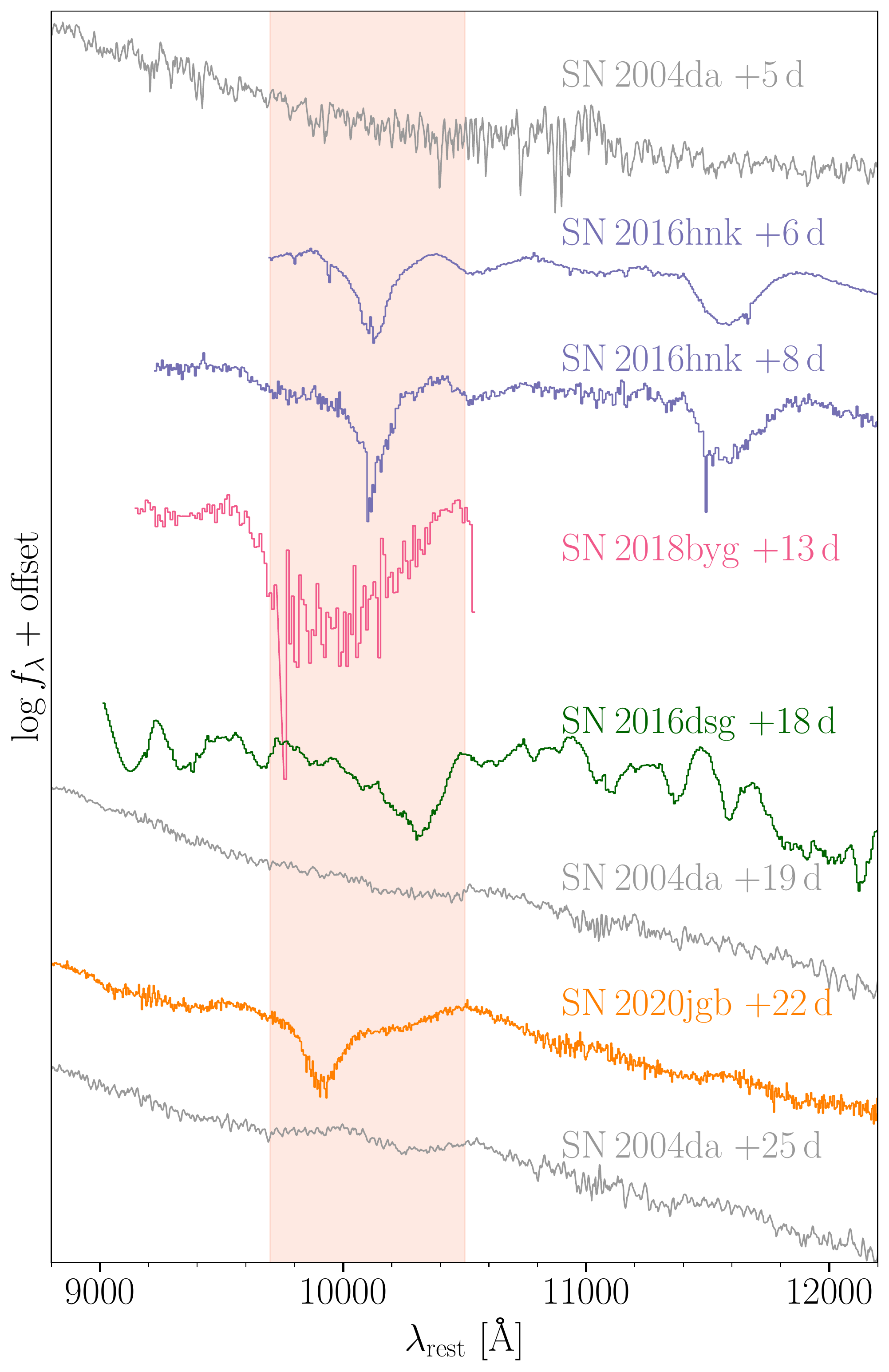}
    \caption{NIR spectra of normal SNe\,Ia SN\,2011fe \citep{Mazzali_2014} and SN\,2004da \citep{Marion2009_NIR} and four subluminous SNe\,Ia as He-shell DDet candidates -- SN\,2016dsg \citep{Dong_16dsg_2022}, SN\,2016hnk \citep{galbany_16hnk_2019}, SN\,2018byg \citep{de_18byg_2019}, and \sn\ (this work). All He-shell DDet candidates show prominent absorption near 1\,\micron\ {(the  highlighted region)}. The spectrum of SN\,2018byg is originally noisy, so it is binned with a size of 10\,\AA. {For SN\,2016dsg, we show the spectrum smoothed with a Savitzky--Golay filter in \citet{Dong_16dsg_2022}.}} 
    \label{fig:NIR_comp}
\end{figure}

\begin{figure*}
    \centering
    \includegraphics[width=\textwidth]{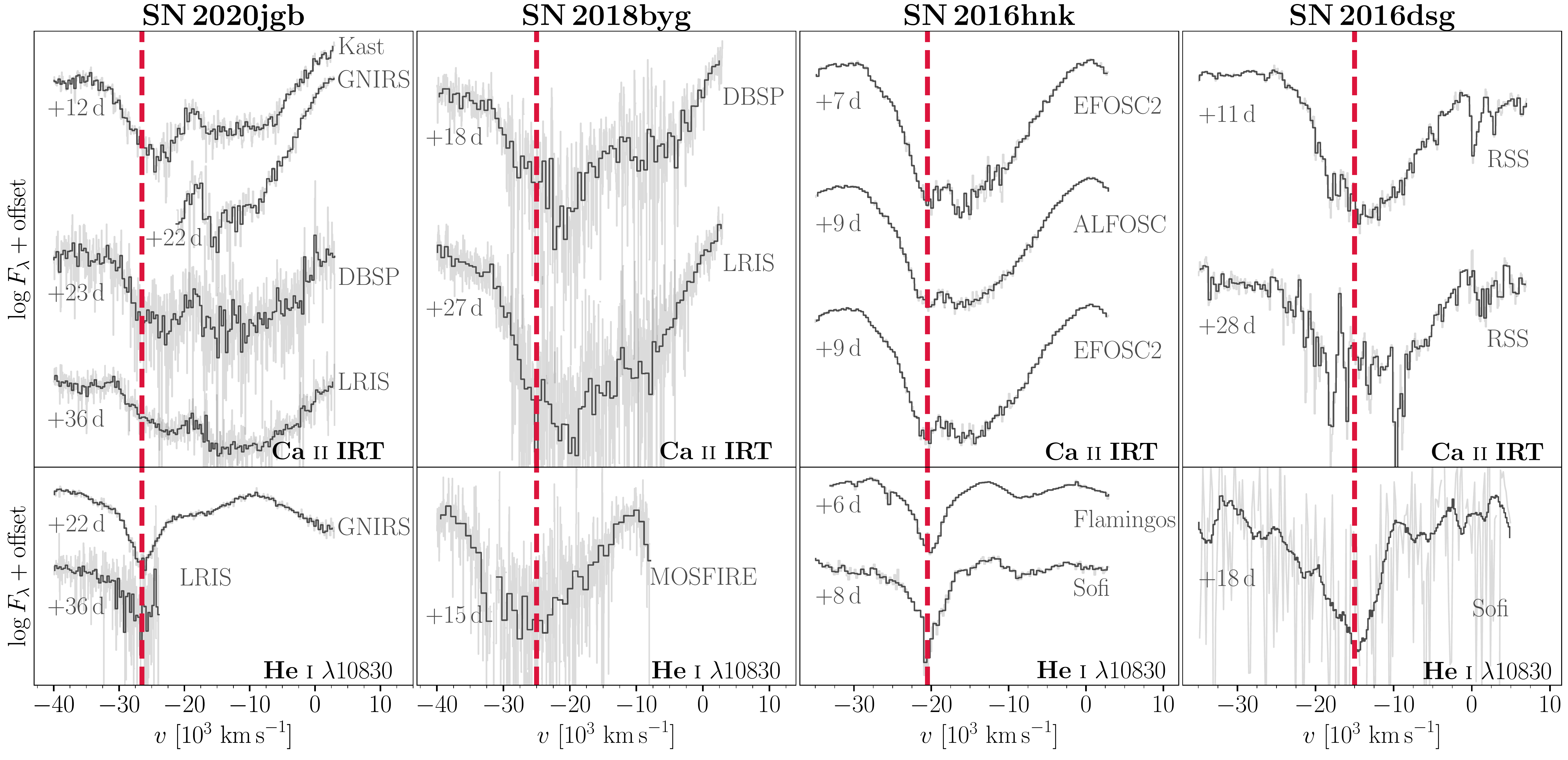}
    \caption{Spectra of \sn, SN\,2018byg \citep{de_18byg_2019}, SN\,2016hnk \citep{galbany_16hnk_2019}, and SN\,2016dsg \citep{Dong_16dsg_2022} in velocity space, showing the similarity in expansion velocities of the 1\,\micron\ features (lower panels) with the \ion{Ca}{2} IRT absorption features (upper panels), assuming the 1\,\micron\ features are associated with \ion{He}{1} $\lambda$10830. The red dashed lines mark the minimum of each 1\,\micron\ feature, which are displayed to guide the eye.}
    \label{fig:hvf_comp}
\end{figure*}

While the nature of the 1\,\micron\ feature remains uncertain, other He-shell DDet candidates show similar complexity in this region. In the currently small sample, only four objects (SN\,2016dsg, SN\,2016hnk, SN\,2018byg, and \sn) have at least one available NIR spectrum (all obtained at different phases), yet each exhibits a strong absorption feature near 1\,\micron, as shown in Figure~\ref{fig:NIR_comp}. SN\,2016hnk has two deep absorption features at $\sim$1.01\,\micron\ and $\sim$1.16\,\micron. It is suggested in \citet{galbany_16hnk_2019} that both of them are caused by \ion{Fe}{2}, though they are deeper than in other SNe\,Ia. {If the 1\,\micron\ feature is associated with \ion{He}{1}, the expansion velocity would be $\sim$21,000\,\kms. For SN\,2016dsg, the minimum of the 1\,\micron\ feature is around $\sim$1.03\,\micron, with a corresponding velocity of $\sim$15,000\,\kms.} The large width and low S/N for the 1\,\micron\ feature in SN\,2018byg make it difficult to determine an exact line velocity, suggesting that feature may be a mixture of several different lines. {Interestingly, all these 1\,\micron\ features, assuming a \ion{He}{1} origin, show an expansion velocity consistent with the HVF of the \ion{Ca}{2} IRT (see Figure~\ref{fig:hvf_comp}). The variation in \ion{Ca}{2} velocities is large ($\sim$15,000--25,000\,\kms), probably due to different viewing angles \citep{Fink_DD_2010,Shen_2D_2021}.} The PVFs of the \ion{Ca}{2} IRT of these He-shell DDet candidates show a similar expansion velocity of $\sim$10,000\,\kms. 

Unfortunately, none of the spectra of SN\,2016dsg, SN\,2016hnk, or SN\,2018byg cover the 2\,\micron\ region; thus, it is not possible to identify the presence of helium decisively. But if the 1\,\micron\ features of these objects are of the same origin, they are more likely to be correlated with the high-velocity ejecta lying in the outmost region in the SNe, because at least for \sn, SN\,2016dsg, and SN\,2016hnk, the difference in their photospheric velocities cannot explain the discrepancy in their line velocities of the 1\,\micron\ feature. Then helium is still a promising candidate to cause strong absorption near 1\,\micron\ for these subluminous He-shell DDet SNe\,Ia.

In conclusion, every peculiar He-shell DDet candidate with available NIR spectra displays a strong absorption feature near 1\,\micron.\footnote{We also note that a similar 1\,\micron\ feature is detected in another possibly relevant object, SN\,2012hn \citep{Valenti_12hn_2014}, in a NIR spectrum obtained at $+25$\,days. SN\,2012hn is a Ca-rich transient exhibiting weak \ion{Si}{2} lines and no optical helium features. 
It shows similar spectral properties (e.g., Fe-group line-blanketing) to those of SN\,2016hnk and SN\,2019ofm \citep{de_Ca_rich_2020}. This indicates a possible He-shell DDet origin of SN\,2012hn.} This feature is not seen in normal SNe\,Ia. Interestingly, the available NIR spectra are all obtained at different epochs, suggesting that this feature may be long-lived. If the feature is due to \ion{He}{1}, then DDet explosions exhibit a wide diversity in the expansion velocity. While it remains to be confirmed in a larger sample, we speculate that anomalously strong absorption around 1\,\micron\ is a distinctive attribute of {peculiar} He-shell DDet SNe.

\subsection{The Host Environment of He-shell DDet SNe} \label{sec:host}
\begin{figure*}
    \centering
    \includegraphics[width=\textwidth]{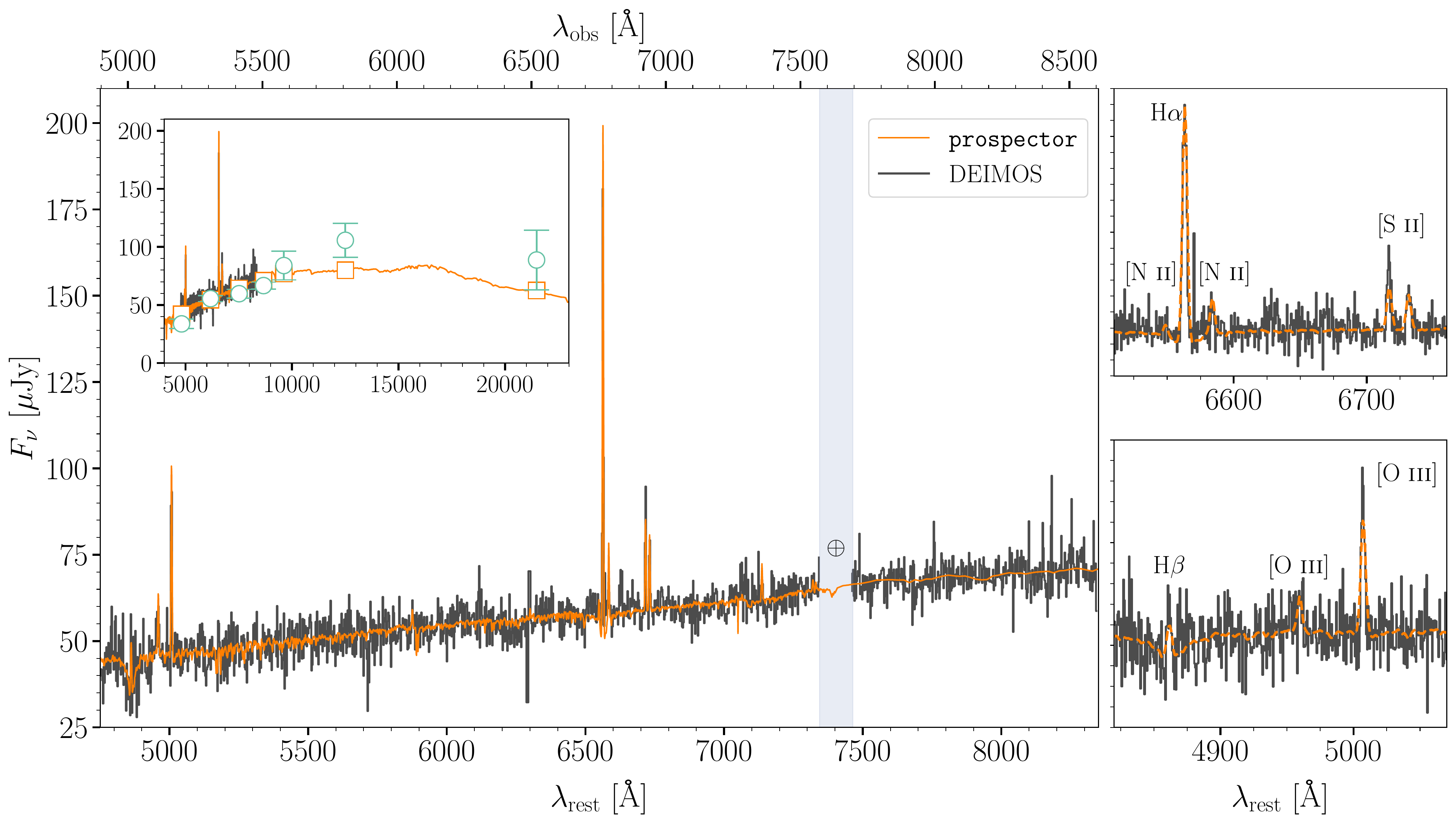}
    \caption{The SED of the star-forming dwarf galaxy PSO J175312.663+005122.078 (the host galaxy of \sn) and the model from \texttt{prospector}. When fitting the SED with \texttt{prospector}, the DEIMOS spectrum is automatically rescaled to fit the archival photometry from Pan-STARRS \citep[][$g$, $r$, $i$, $z$, $y$ Kron magnitudes]{PS1_2016} and VHS \citep[][$J$ and $K_s$ Petrosian magnitudes]{VHS_2013}. {\it Left:} the SED in the optical band (4750--8350\,\AA\ in the rest frame of the host galaxy). The black line corresponds to the observed spectrum, binned with a size of 2\,\AA. The orange line is the \texttt{prospector} model produced from the median of posterior distributions of the stellar population properties. The blue shaded region is masked in the fitting owing to the strong telluric lines. The inset shows the same comparison, but covering the $g$ through $K_s$ bands (4000--24,000\,\AA). Apart from the spectra, we also show the multiband photometry (green circles) and the best-fit magnitudes (orange squares). {\it Right:} spectra around the most prominent emission lines. {\it Top right:} H$\alpha$, [\ion{N}{2}] $\lambda\lambda$6548, 6583, [\ion{S}{2}] $\lambda\lambda$6716, 6731. {\it Bottom right:} H$\beta$, [\ion{O}{3}] $\lambda\lambda$4959, 5007.}
    \label{fig:host_spec}
\end{figure*}

\begin{figure}
    \centering
    \includegraphics[width=\linewidth]{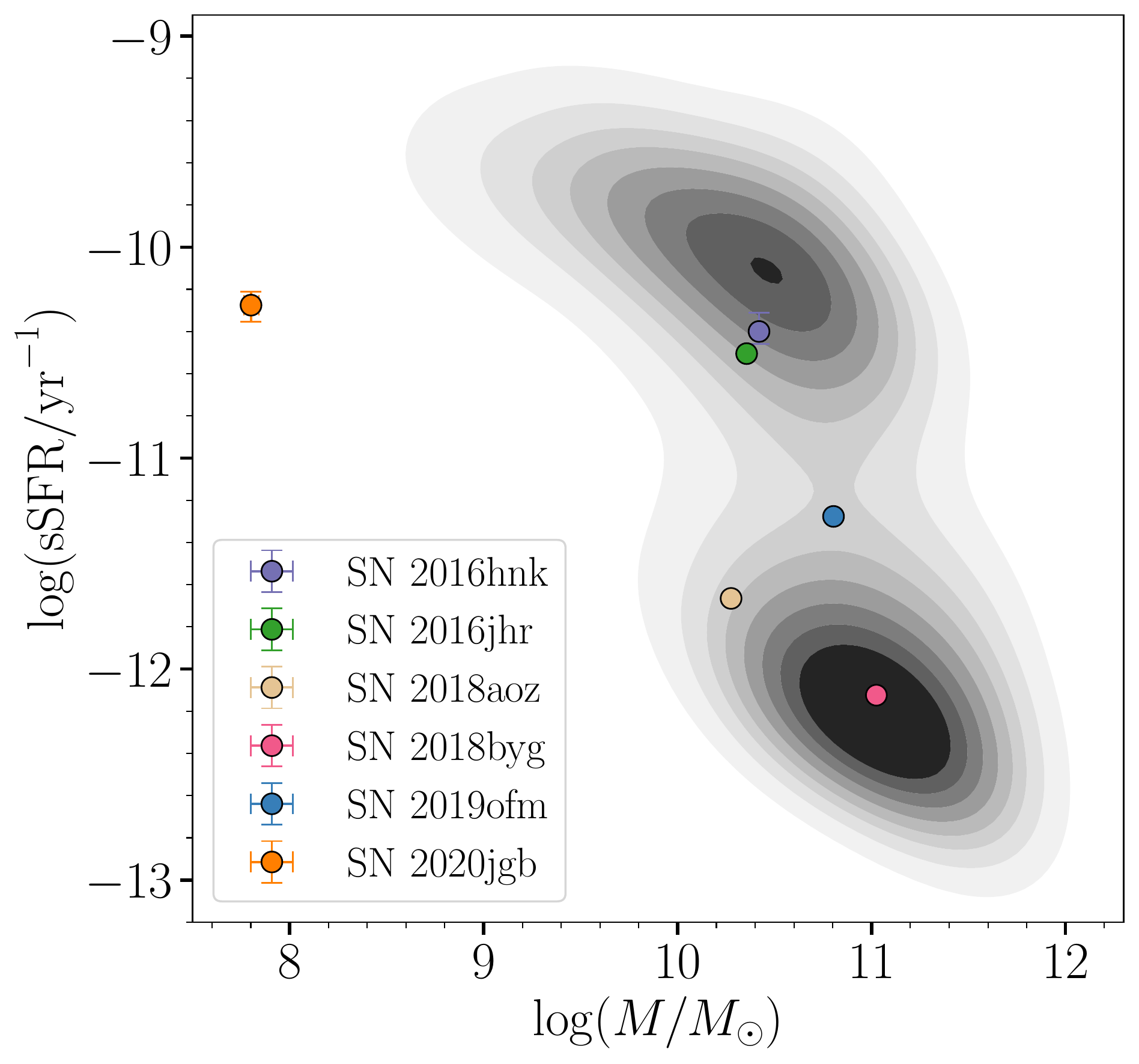}
    \caption{The sSFR and stellar mass for the host galaxies of He-shell DDet candidates, showing that He-shell DDet SNe can emerge in both star-forming and passive galaxies. The properties for the hosts of SN\,2016hnk and SN\,2018aoz are taken from \citet{Dong_Ca-rich_2022} and the CLU catalog \citep{Cook_2019, de_Ca_rich_2020}, respectively. The gray contours correspond to the bivariate distributions of stellar mass and sSFR for galaxies in the SDSS MPA-JHU DR8 catalog \citep{Kauffmann_SDSS_2003,Brinchmann_SDSS_2004}, visualized using kernel density estimation with the data visualization library \texttt{seaborn} \citep{Waskom_seaborn_2021}. Galaxies with BPT classification as AGNs or LINERs are excluded, since certain spectral features (e.g., H$\alpha$ emission) due to nuclear activity might be misinterpreted as being caused by star formation.}
    \label{fig:host}
\end{figure}

We model the host galaxy of \sn\ (see Figure~\ref{fig:host_spec}) using \texttt{prospector} \citep{Johnson_prospector_2021}, a package for principled inference of stellar population properties using photometric and/or spectroscopic data. \texttt{Prospector} applies a nested sampling fitting routine through \texttt{dynesty} \citep{Speagle_dynesty_2020} to the observed data and produces posterior distributions of the stellar population properties and model SEDs with use of \texttt{Python-FSPS} \citep{Conroy_2009,Conroy_2010}. Our observed data include the Galactic-extinction-corrected DEIMOS spectrum, as well as the archival photometric data from the Panoramic Survey Telescope and Rapid Response System \citep[Pan-STARRS;][$g$, $r$, $i$, $z$, $y$ Kron magnitudes]{PS1_2016}  and the VISTA Hemisphere Survey \citep[VHS;][$J$ and $K_s$ Petrosian magnitudes]{VHS_2013}. We use a parametric delayed-$\tau$ star-formation history, given by Equation~(1) of \citet{Nugent_2020} and defined by the $e$-folding factor $\tau$, the Galactic dust extinction law \citep{Cardelli_1989}, and the Chabrier initial mass function \citep{Chabrier_2003} to the model. We further apply a mass-metallicity relation \citep{Gallazzi_2005} to sample realistic stellar masses and metallicities and a dust law that ensures young stellar light attenuates dust twice as much as old stellar light, as has been observed.  We also add a nebular emission model \citep{Byler_2017} with a gas-phase metallicity and a gas ionization parameter to correctly measure the strength of the emission lines in the DEIMOS spectrum. The model spectral continuum is built from a tenth-order Chebyshev polynomial. We determine the stellar mass and star-formation rate (SFR) from the \texttt{prospector} output, as shown by \citet{Nugent_2022}. The estimated stellar mass is $\log (M_*\,[\Msun])=7.79_{-0.06}^{+0.07}$, and the specific star-formation rate (sSFR) is $\log (\mathrm{sSFR}\,[\mathrm{yr}^{-1}])=-10.25_{-0.08}^{+0.09}$, with the uncertainties denoting the 68\% highest posterior density regions.
  
In Figure~\ref{fig:host}, we show the sSFR and the stellar mass for the host galaxies of six He-shell DDet candidates. Again using \texttt{prospector}, we fit the stellar properties for all the other candidates with optical spectra from the Sloan Digital Sky Survey \citep[SDSS;][]{York_2000} and photometry from the DESI Legacy Imaging Surveys \citep[LS;][$g$, $r$, $z$, $W_1$, $W_2$, $W_3$, $W_4$ magnitudes]{Dey_2019}. With mid-infrared (MIR) photometry\footnote{LS DR9 includes MIR ($W_1$--$W_4$) fluxes from images through year 6 of NEOWISE-Reactivation (\url{https://wise2.ipac.caltech.edu/docs/release/neowise/neowise_2020_release_intro.html}) force-photometered in the unWISE \citep{Meisner2017a,Meisner2017b,unwise_images} maps at the locations of LS optical sources.} available, \texttt{prospector} can better estimate the overall dust extinction in the host galaxy and the contribution of an active galactic nucleus (AGN) to the SED. We therefore add two additional parameters to our \texttt{prospector} fit to sample the MIR optical depth and fraction of the total galaxy luminosity due to an AGN.

Unfortunately, two hosts (those of SN\,2016hnk and SN\,2019ofm) are nearby ($z \lesssim 0.03$) late-type galaxies with extended, spatially resolved spiral structures. Examination of the photometry model from Legacy Surveys (LS) shows that the galaxy aperture does not include the blue, diffuse star-forming regions of these galaxies. Fitting the SDSS spectra + LS photometry would inevitably underestimate their sSFR. For the host of SN\,2016hnk, we instead adopt the results of \citet{Dong_Ca-rich_2022}, which are based on broadband far-ultraviolet to far-infrared photometry from the $z=0$ Multiwavelength Galaxy Synthesis I \citep[z0MGS;][]{Leroy_2019} to characterize the stellar population with \texttt{prospector}. The SFR they estimated is 1.1\,dex higher than ours, suggesting intense star-formation in the spiral arms.
For the host of SN\,2019ofm, there are no archival stellar population data available; so we redo the photometry using science-ready coadded images from the \textit{Galaxy Evolution Explorer} (GALEX) general release 6/7 \citep[][$FUV$ and $NUV$ bands]{Martin2005a}, SDSS DR9 (\citealp{Ahn2012a}, $u$, $g$, $r$, $i$, $z$ bands), the Two Micron All Sky Survey \citep[2MASS;][$H$ and $J$ bands]{2mass,Skrutskie2006a}, and preprocessed WISE images \citep{Wright2010a} from the unWISE archive \citep[][$W_1$ and $W_2$ bands]{Lang2014a}.\footnote{The unWISE images are based on the public WISE data and include images from the ongoing NEOWISE-Reactivation mission R3 \citep{Mainzer2014a, Meisner2017a}, available on \href{http://unwise.me}{http://unwise.me}.} We use the software package \texttt{LAMBDAR} (Lambda Adaptive Multi-Band Deblending Algorithm in R) \citep{Wright2016a} and tools presented in \citet{Schulze2021a}, to measure the total brightness of the host galaxy. But with the \texttt{LAMBDAR} photometry, the estimated SFR is essentially the same as in the previous fit, suggesting that there is not much ongoing star-formation in the spiral arms. This, along with its moderate sSFR ($\log (\mathrm{sSFR}\,[\mathrm{yr}^{-1}])=-11.27$), indicates the host galaxy is in the transitional phase.

In addition, the host of the normal SN\,Ia SN\,2018aoz (NGC\,3923) is a local ($z=0.00580$) early-type galaxy and is outside the SDSS footprint, so we adopt its stellar population properties from the Census of the Local Universe (CLU) catalog \citep{Cook_2019, de_Ca_rich_2020}. Nonetheless, it is close to several extended sources with low surface brightness, which could be faint dwarf galaxies \citep[see Figure~3 in][]{Kasliwal_2012}. Its nebular-phase spectrum exhibits H$\alpha$ emission, which indicates potential star formation, but could also be explained with photoionized gas around the transient \citep{Kasliwal_2012}.

Figure~\ref{fig:host} reveals that He-shell DDet SNe emerge in both star-forming and passive galaxies. There is also significant diversity in their location within their host galaxy. SN\,2020jgb has a small projected physical offset ($\sim$0.2\,kpc) from the center of its host, a star-forming dwarf galaxy, so it is likely to originate from a young, star-forming environment. SN\,2016hnk has a moderate projected host offset ($\sim$4\,kpc) and a potential origin in an \ion{H}{2} region with ongoing star formation \citep{galbany_16hnk_2019}. SN\,2019ofm has a large projected offset ($\sim$11\,kpc) but is still on a spiral arm, as shown in its DECaLS image \citep{Dey_2019}. Other objects, including the recently reported SN\,2016dsg and OGLE-2013-SN-079 \citep{Dong_16dsg_2022}, show large projected host offsets ($\gtrsim$10\,kpc) and lie in the galaxy outskirts, which usually indicates an old stellar population origin.

In this sense, the He-shell DDet sample resembles the normal SN\,Ia population, which can occur in both star-forming and quenched galaxies \citep[e.g.,][]{Sullivan_2006, Smith_2012}. This is very different from some other types of thermonuclear SNe such as 02cx-like SNe, which almost only appear in star-forming galaxies, or 91bg-like and 02es-like objects, which prefer old stellar environments \citep[see the review by][]{Jha_2019}. This favors the postulated sequence that He-shell DDet SNe may make up a substantial fraction of normal SNe\,Ia, and is supported by observations of stellar metallicity \citep{Sanders_2021, Eitner_2022}.

The diversities in host environments indicate multiple formation channels in the He-shell DDet SN population. Those in star-forming galaxies, \sn\ being the most unambiguous example, could originate from some analogues of the two subdwarf B binaries with WD companions \citep{Iben_1987,Geier_2013, Kupfer_2022} discovered in young stellar populations.
On the other hand, those with large host offsets could not be easily formed {\it in situ}. Similarly, many Ca-rich transients \citep{Filippenko_2003, Perets_2010, Kasliwal_2012} are also observed in remote locations \citep[e.g.,][]{Lunnan_2017}, for which some dynamical formation channels have been proposed \citep{Lyman_2014}. To reach the outskirts of galaxies, WD binaries would need to be ejected by globular clusters \citep{Shen_2019} or supermassive black holes \citep{Foley_2015} before explosion. Given that some Ca-rich transients show characteristic DDet properties \citep{de_Ca_rich_2020}, these channels may also be applicable to some of the He-shell DDet SNe. 

The robust detection of \sn\ in a star-forming region also agrees with independent studies of SN\,Ia progenitors using observations of stellar metallicity. After measuring the manganese abundance in the Sculptor dwarf spheroidal galaxy, it is argued in \citet{de_los_reyes_manganese_2020} that sub-\Mch\ SNe\,Ia dominate the initial chemical enrichment of a galaxy, while near-\Mch\ SNe become more important at later times. This indicates that, observationally, sub-\Mch\ SNe\,Ia might have a stronger preference toward younger stellar populations than near-\Mch\ SNe\,Ia. 
We note that while \sn\ is the first confirmed subluminous He-shell DDet SN in a star-forming dwarf, which indicates that peculiar He-shell DDet SNe might be intrinsically rare, the same may not be true for {the potential population of normal SNe\,Ia ignited by a DDet \citep{Magee_2021}}. {A red flux excess, the hallmark of an He-shell detonation in a normal SN\,Ia, will only be evident in the first few days after the explosion, while} few SNe\,Ia have been observed at such an early phase to date; thus, we might have missed a great number of {normal} He-shell DDet SNe. A systematic study based on prompt follow-up observations of infant SNe\,Ia will help verify this implication.

\section{Conclusions} \label{sec:conclusion}
We have presented observations of \sn, a peculiar SN\,Ia. It has a low luminosity, red $g_\mathrm{ZTF}-r_\mathrm{ZTF}$ colors, and strong line-blanketing in the optical spectra near maximum light. These observational properties are very similar to those of SN\,2018byg \citep{de_18byg_2019}, which could be explained by the detonation of a shell of helium on a sub-\Mch WD. Fitting the light curves of \sn\ to a grid of models from \citet{polin_observational_2019}, we show that a $\sim$0.82\,$\Msun$ WD beneath a $\sim$0.13\,$\Msun$ He-shell provides a reasonable match to the peak-time spectrophotometric properties of \sn. {The systematics in our radiative transfer models, however, result in significant uncertainty in the shell mass.} The uncertainty in the host-galaxy extinction also limits the precision on estimating total progenitor mass, with a reasonable upper limit being $\sim$1.00\,$\Msun$. 

A high-S/N NIR spectrum obtained three weeks after maximum light exhibits a prominent absorption feature near 1\,\micron, which could be produced by the unburnt helium (\ion{He}{1} $\lambda$10830) in the outermost ejecta expanding at a high velocity ($\sim$26,000\,\kms). At the same epoch, the \ion{Ca}{2} IRT also has similarly high velocities ($\sim$24,000\,\kms). To date, NIR spectra have been observed for only a handful of candidate He-shell DDet SNe. Interestingly, all of them show deep absorption features near 1\,\micron, which, if assumed to be \ion{He}{1} $\lambda$10830, would be expanding at a very similar velocity to the HVFs of the \ion{Ca}{2} IRT. For these candidates the \ion{Ca}{2} HVFs and putative \ion{He}{1} velocities show significant diversity, ranging from $\sim$15,000\,\kms\ in SN\,2016dsg to $\sim$24,000\,\kms\ in \sn. If it is the unburnt helium and the newly synthesized calcium from the He-shell that produce these line features, such a consistency in the expansion rates of different absorption lines would be naturally explained. However, we could not find unambiguous evidence for other \ion{He}{1} absorption lines, such as \ion{He}{1} $\lambda$20581, so we cannot claim a definitive detection of helium in \sn. Nonetheless, alternative possibilities (\ion{Mg}{2}, \ion{C}{1}, \ion{Fe}{2}) that may cause the 1\,\micron\ feature are deemed even less likely. Helium is thus the most plausible explanation for the apparently ubiquitous 1\,\micron\ features.

We propose that He-shell DDet SNe can be robustly identified with NIR spectra. For transients showing a clear 1\,\micron\ feature, its potential association with \ion{He}{1} $\lambda$10830 could be tested by following the checklist below.
\begin{itemize}
    \item Search for \ion{He}{1} $\lambda$20581. A caveat is that one should not always expect to see significant \ion{He}{1} $\lambda$20581 absorption in He-shell DDet SNe, since this line is weaker than \ion{He}{1} $\lambda$10830 and could be almost invisible when the He-shell is thin \citep{Boyle2017_Helium}. Strong telluric lines near 2\,\micron\ can make it difficult to detect \ion{He}{1} $\lambda$20581.
    \item Calculate the line velocity assuming the feature is \ion{He}{1} $\lambda$10830 and check whether the speed is comparable with the \ion{Ca}{2} IRT HVFs at a similar phase. While both the detonation recipe and viewing angles would affect the observed \ion{He}{1}/\ion{Ca}{2} velocity, we still expect both elements to expand at similar speeds along the line of sight if they both have an He-shell origin.
    \item Exclude the possibility of other strong lines. If the NIR spectrum is obtained before the peak brightness of the SN, strong \ion{Mg}{2} and \ion{C}{1} absorption \citep{Hsiao_CSP_2019} would be possible contaminants. Otherwise, if the 1\,\micron\ feature is seen in the transitional-phase spectrum when the inner region of the SN becomes visible, we need to carefully rule out the possibility of an \ion{Fe}{2} origin \citep{Marion2009_NIR}.
\end{itemize}

The small, but growing, sample of He-shell DDet SNe are heterogeneous in their observational properties, including peak luminosity, color evolution, chemical abundances, and line velocities, which could be explained by a large variety of He-shell and WD masses \citep{polin_observational_2019,Shen_2D_2021}, viewing angles \citep{Shen_2D_2021}, and the initial chemical compositions in the He-shell \citep{Kromer_DD_2010}. In addition, they are discovered in both old and young stellar populations, \sn\ being the first unambiguous {peculiar} He-shell DDet candidate in a star-forming dwarf galaxy. If, as has been argued \citep[e.g.,][]{Sanders_2021, Eitner_2022}, a substantial fraction of normal SNe\,Ia are triggered by He-shell DDets, then we would naturally expect He-shell DDet SNe to emerge in both star-forming and passive galaxies. Our discovery of \sn\ in a star-forming dwarf galaxy confirms that He-shell DDet events occur in a variety of different galaxies. This is unlike some other subtypes of SNe\,Ia \citep{Jha_2019}, which strongly prefer either star-forming galaxies (e.g., SNe\,Iax) or passive galaxies (e.g., 91bg-like and 02es-like objects). Nonetheless, it remains to be examined whether {peculiar} He-shell DDet SNe stem from similar progenitors to the normal SNe\,Ia triggered by {a DDet}, or whether their massive He-shells could only be developed in a completely distinctive population of binary systems.\\


\noindent {We thank the anonymous referee for a thoughtful and detailed report.} We thank Eddie Schlafly and Dustin Lang for suggesting photometry from DESI Legacy Imaging Surveys in SED fitting. We are grateful to Aishwarya Dahiwale, Jillian Rastinejad, and Yuhan Yao for the high-quality spectra they obtained. We also appreciate the excellent assistance of the staffs of the various observatories where data were obtained. K.D. acknowledges support from NASA through the NASA Hubble Fellowship grant \#HST-HF2-51477.001 awarded by the Space Telescope Science Institute, which is operated by the Association of Universities for Research in Astronomy, Inc., for NASA, under contract NAS5-26555. A.V.F. is grateful for financial support from the Christopher R. Redlich Fund and many other individual donors. K.M. is funded by the EU H2020 ERC grant No. 758638. S.S. acknowledges support from the G.R.E.A.T research environment, funded by {\em Vetenskapsr\aa det}, the Swedish Research Council, project number 2016-06012. This work was also supported by the GROWTH project \citep{Kasliwal2019} funded by the National Science Foundation (NSF) under grant 1545949.

This work is based on observations obtained with the Samuel Oschin Telescope 48-inch and the 60-inch Telescope at the Palomar Observatory as part of the Zwicky Transient Facility project. ZTF is supported by the National Science Foundation under Grant No. AST-1440341 and a collaboration including Caltech, IPAC, the Weizmann Institute of Science, the Oskar Klein Center at Stockholm University, the University of Maryland, the University of Washington, Deutsches Elektronen-Synchrotron and Humboldt University, Los Alamos National Laboratories, the TANGO Consortium of Taiwan, the University of Wisconsin at Milwaukee, and Lawrence Berkeley National Laboratories. Operations are conducted by COO, IPAC, and UW. 
SED Machine is based upon work supported by the National Science Foundation under Grant No.\ 1106171.

This work is also based on observations made with the Nordic Optical Telescope, owned in collaboration by the University of Turku and Aarhus University, and operated jointly by Aarhus University, the University of Turku and the University of Oslo, representing Denmark, Finland and Norway, the University of Iceland and Stockholm University at the Observatorio del Roque de los Muchachos, La Palma, Spain, of the Instituto de Astrofisica de Canarias.

A major upgrade of the Kast spectrograph on the Shane 3\,m telescope at Lick Observatory, led by Brad Holden, was made possible through gifts from the Heising-Simons Foundation, William and Marina Kast, and the University of California Observatories. Research at Lick Observatory is partially supported by a generous gift from Google. The W. M. Keck Observatory is operated as a scientific partnership among the California Institute of Technology, the University of California and NASA; the observatory was made possible by the generous financial support of the W. M. Keck Foundation. {W. M. Keck Observatory access was supported by Northwestern University and the Center for Interdisciplinary Exploration and Research in Astrophysics (CIERA).}


\facility{PO:1.2m (ZTF), PO:1.5m (SEDM), Gemini:Gillett (GNIRS), Hale (DBSP), NOT (ALFOSC), Shane (Kast Double spectrograph), Keck:I (LRIS), Keck:II (DEIMOS).}
\software{\texttt{astropy} \citep{Astropy_2013, Astropy_2018}, \texttt{CASTRO} \citep{Almgren_Castro_2010}, \texttt{dynesty} \citep{Speagle_dynesty_2020}, \texttt{emcee} \citep{emcee_2013}, \texttt{LAMBDAR} \citep{Wright2016a}, \texttt{matplotlib} \citep{Matplotlib_2007}, \texttt{prospector} \citep{Johnson_prospector_2021}, \texttt{PypeIt} \citep{pypeit:zenodo}, \texttt{pysedm} \citep{Rigault_pysedm_2019}, \texttt{Python-FSPS} \citep{Conroy_2009,Conroy_2010}, \texttt{scipy} \citep{Scipy_2020}, \texttt{seaborn} \citep{Waskom_seaborn_2021}, \texttt{SEDONA} \citep{Kasen_Sedona_2006}.}

\appendix
\begin{figure*}
    \centering
    \includegraphics[width=\textwidth]{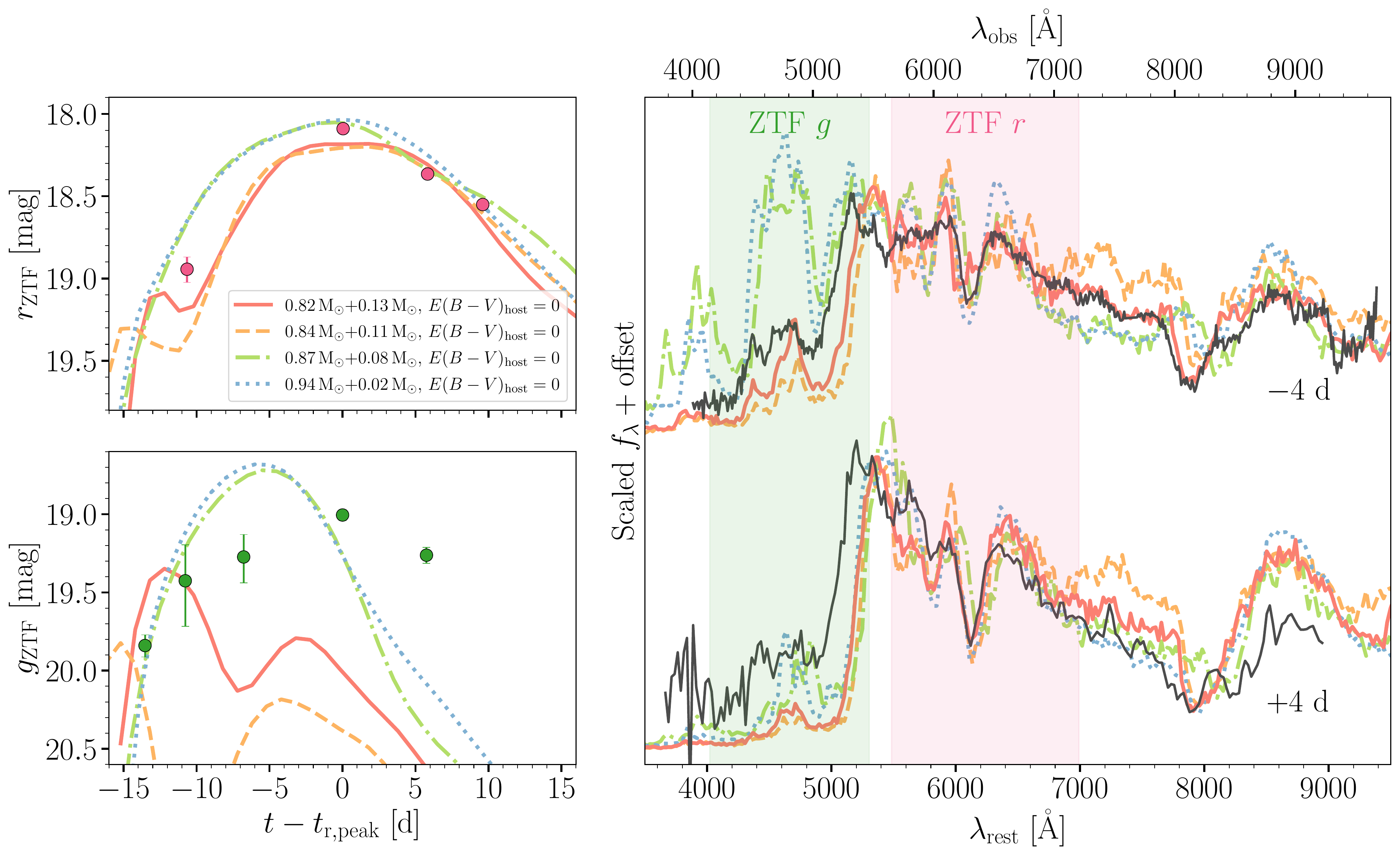}
    \caption{Similar to Figure~\ref{fig:model}, but more models with a total mass of $\sim$0.95\,$\Msun$ and various shell masses (from $0.02\,\Msun$ to $0.13\,\Msun$) are displayed. For these models, we assume no host extinction.}
    \label{fig:model_0_95}
\end{figure*}

\begin{figure*}
    \centering
    \includegraphics[width=\textwidth]{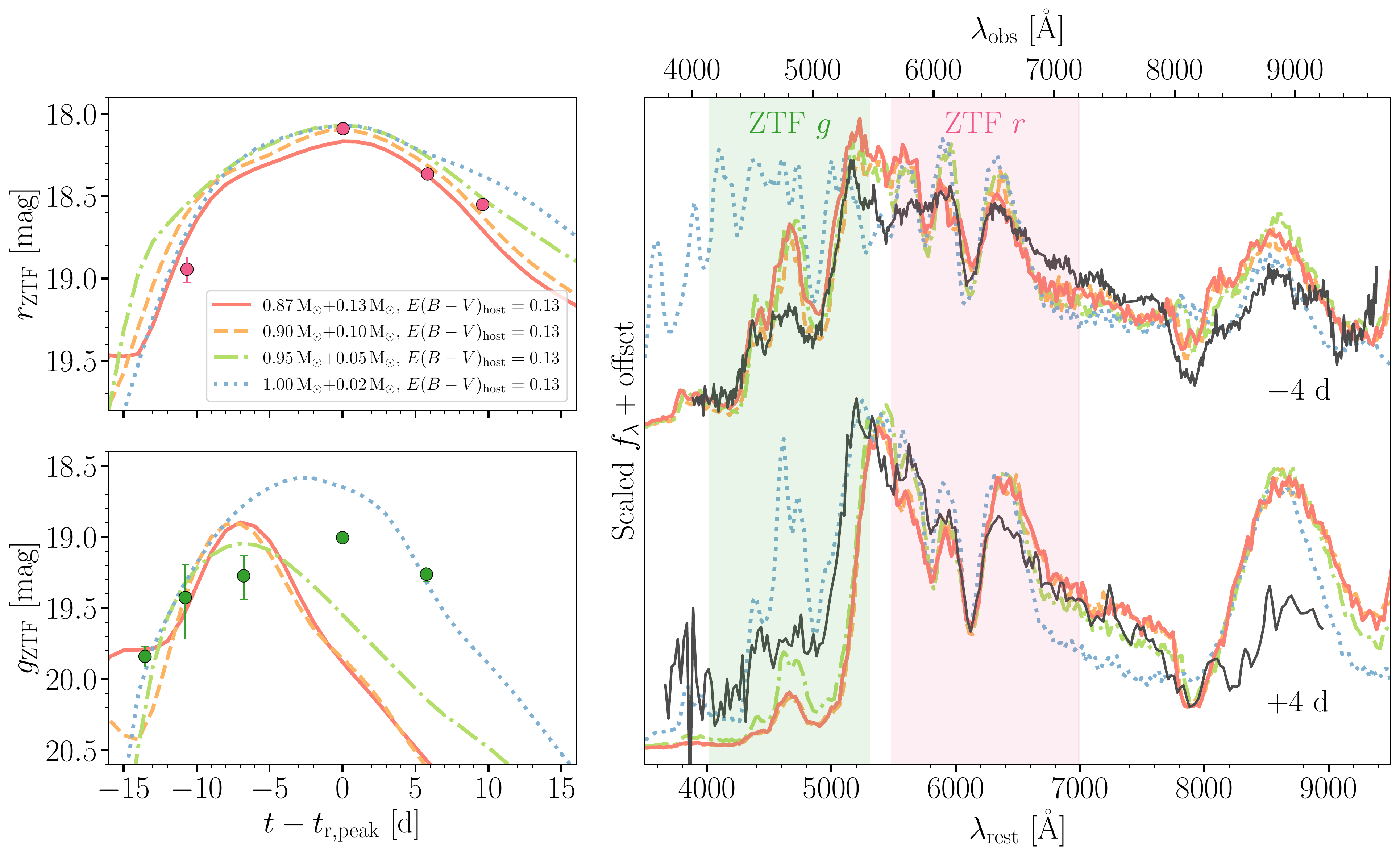}
    \caption{Similar to Figure~\ref{fig:model}, but more models with a total mass of $\sim$1.00\,$\Msun$ and various shell masses (from $0.02\,\Msun$ to $0.13\,\Msun$) are displayed. For these models, we assume $E(B-V)_\mathrm{host}=0.13$.}
    \label{fig:model_1_0}
\end{figure*}

\section{Comparison to DDet Models with Various Shell Masses}\label{app1}
We have shown that the $r_\mathrm{ZTF}$-band light curve and the observed spectra of \sn\ near maximum brightness are fairly consistent with the DDet of a sub-\Mch\ WD beneath a massive shell ($\sim$0.13\,$\Msun$), whose total mass is $\sim$0.95--1.00\,$\Msun$. In this appendix we compare \sn\ to other DDet models developed using the methods in \citet{polin_observational_2019}. 

Figure~\ref{fig:model_0_95} shows multiple models with a total mass of $\sim$0.95\,$\Msun$, all of which reproduce the brightness of \sn\ in $r_\mathrm{ZTF}$ if there is no host extinction. The $g_\mathrm{ZTF}$-band synthetic light curves, which depend on the strength of line-blanketing of the Fe-group elements, differ significantly depending on the He-shell mass. In the two models with thinner shells ($\lesssim$0.08\,$\Msun$), the suppression of flux blueward of $\sim$5000\,\AA\ is much less significant than that seen in \sn\ at $-$4\,days. As a result, they overestimate the brightness in $g_\mathrm{ZTF}$ before maximum light. The $0.84\,\Msun+0.11\,\Msun$ model shows the most significant line-blanketing.

Figure~\ref{fig:model_1_0} shows models with a total mass of $\sim$1.00\,$\Msun$ assuming $E(B-V)_\mathrm{host}=0.13$\,mag. Each model reproduces the brightness of \sn\ in $r_\mathrm{ZTF}$. The model with the thinnest shell significantly underestimates the level of line-blanketing, allowing us to eliminate it as a viable model for \sn. Models with shells $\gtrsim$0.05\,$\Msun$ exhibit similar behavior, meaning the shell mass is quite uncertain.
We note that all the 1.00\,$\Msun$-models overestimate the maximum brightness of \sn\ in $g_\mathrm{ZTF}$ and underestimate the level of line-blanketing in the spectrum at $-4$\,days.

While none of the models presented here provides a perfect match to the observations, {\sn\ is more consistent with He-shell DDet models with relatively massive ($\gtrsim$0.1\,$\Msun$) shells}.

\bibliography{SN2020jgb, ZTF, software}
\bibliographystyle{aasjournal}



\end{CJK*}
\end{document}